\documentclass[fleqn,twoside]{article}

\usepackage{epsf,multicol,ifthen}

\usepackage{hep}

\usepackage[cp1251]{inputenc}
\usepackage[english,ukrainian,russian]{babel}
\usepackage{graphicx,amsmath,amssymb}

\nazva{Annihilation Mechanism of Dilepton Emission from Finite Fireball}

\font\Bbb=msbm10 scaled \magstep1 \font\Bbbscr=msbm7 scaled
\newfam\Bbbfam
\textfont\Bbbfam=\Bbb \scriptfont\Bbbfam=\Bbbscr
\scriptscriptfont\Bbbfam=\Bbbscrscr
\def\Bbb{\Bbbfam}

\font\Cal=msbm10 scaled \magstep1 \font\Calscr=msbm7 scaled
\newfam\Calfam
\textfont\Calfam=\Cal \scriptfont\Calfam=\Calscr
\scriptscriptfont\Calfam=\Calscrscr
\def\Cal{\fam\Calfam}
\def\beq{\begin{equation}}
\def\eeq{\end{equation}}

\def\gappeq{\mathrel{\rlap {\raise.5ex\hbox{$>$}}
{\lower.5ex\hbox{$\sim$}}}}

\def\lappeq{\mathrel{\rlap{\raise.5ex\hbox{$<$}}
{\lower.5ex\hbox{$\sim$}}}}

\newcommand{\lsim}{\raise.3ex\hbox{$<$\kern-.75em\lower1ex\hbox{$\sim$}}}
\newcommand{\gsim}{\raise.3ex\hbox{$>$\kern-.75em\lower1ex\hbox{$\sim$}}}

\udk{04-25E}
\nazvacol{ Annihilation Mechanism of Dilepton  }%
\avtor{D. V. ANCHISHKIN, V. M. KHRYAPA, R. A. NARYSHKIN$^1$, P.~V.~RUUSKANEN$^2$ }%
\avtorcol{D. V. ANCHISHKIN, V. M. KHRYAPA, R. A. NARYSHKIN,  P.~V.~RUUSKANEN }%
\inst{Bogolyubov Institute for Theoretical Physics, Nat. Acad. of
Sci. of Ukraine}%
\adr{(14b Metrolohichna Str., 03143 Kyiv, Ukraine)}%
\insti{Taras Shevchenko Kiev National University, Physics
Department} \adri{(6 Glushkova Prosp., 03022 Kiev, Ukraine)}
\instii{University of Jyv\"askyl\"a, Physics Department}%
\adrii{(P.O.Box 35, FIN-40351 Jyv\"askyl\"a, Finland)}%
\begin{document}
\setcounter{page}{001}%

\maketitl

\begin{multicols}{2}

\anot{Medium-induced modifications of the $\pi^+\pi^-$ and
$q\overline{q}$ annihilation mechanisms of dilepton production during
relativistic heavy ion collisions are considered. Due to the dense
hadron environment, the pions produced during a collision are
effectively confined in a finite volume, in which they live for a
finite time which is scaled as the lifetime of a fireball. Keeping
the vacuum mass and width of the $\rho$-meson form-factor, we
compare two approaches to the description of the $\rho$-meson
behavior. In the first approach, a $\rho$-meson has a zero mean
free path due to the dense hadron environment. But, in the second
approach, it propagates as in vacuum. Our results indicate that,
due to the space-time finiteness of the pion system which
generates the corresponding quantum randomization, the dilepton
rates are finite in the invariant low-mass region $M\leqslant
2m_\pi$. It is found that the spatial finiteness of quark wave
functions and the finiteness of the lifetime of excited states
result in the same effect for the $q\overline{q}$ annihilation to
dileptons. The breaking of the detailed energy-momentum
conservation due to the broken translation invariance 
is discussed. }

\section*{Introduction}

Lepton pairs produced in high energy nucleus-nucleus collisions
provide an information on the early high-temperature and
high-density stage, where the quark-gluon plasma (QGP) formation
is expected. The registration of dileptons gives important
observables which probe the the pion dynamics in the dense nuclear
matter that exists on the early stage of the collision. After the
creation, $e^+\, e^-$ and $\mu^+\, \mu^-$ pairs do not practically
interact with the surrounding hadron matter, and the analysis of
the final dilepton spectra experimentally obtained provides an
excellent possibility to investigate the expansion dynamics of
hadron fireballs. The enhancement in the production of dileptons
with an invariant mass of 200-800 MeV observed by the CERES
collaboration \cite{agakichev95,agakichev98} has received a
considerable attention recently and has been studied in the
framework of various theoretical models. It was found  that the
most prominent channel of the dilepton production accounting a
large part of the observed enhancement is the pion annihilation
\cite{koch96,haglin}. The sensitivity of the dilepton spectra to a
variation in the initial conditions for a hadron fireball was
investigated. Namely, an enhancement of the dilepton production
was explained  by a modification of pions in the medium, the
pion-nuclear form-factors in the pion gas \cite{koch96}, or the
pion dispersion relation modified in the medium
\cite{rapp96,song}. The inclusion of the effects of baryon
resonances \cite{rapp97,rapp99} which couple directly to
$\rho$-mesons seems to be able to substantially increase the
yield, although the calculation in \cite{steele} found a much
smaller effect due to baryons. In \cite{kluger}, the dilepton
production was calculated as a result of the interaction between
long-wavelength pion oscillations or disoriented chiral
condensates and the thermal environment within the linear sigma
model. It was shown that the dilepton yield with the invariant
mass near and below $2m _\pi$ can be larger up to two orders of
magnitude than the corresponding equilibrium yield due to soft
pion modes. Finite pion width effects and their influence on
$\rho$-meson properties and, subsequently, on dilepton spectra
were investigated in \cite{van-hees2000}, and the influence of
nonequilibrium processes and finite times were investigated in
\cite{cooper}. The conclusion of many works is that, in order to
better fit the experimental data, an additional modifications,
which are due to the medium and nonequilibrium processes, should
be considered (for recent review, see \cite{gale-haglin-2003}).

The purpose of the present paper is to revise the annihilation
mechanism of dilepton production by taking into account that a
multipion system or a multiquark system is confined to a finite
space-time region. The standard consideration of the reaction $
\pi^+\, \pi^- \to  \rho \to \gamma^* \to l\, \overline{l} $ or $
q\, \overline{q} \to l\, \overline{l} $
\cite{domokos,mclerran85,gale87,wong} assumes an infinite
space-time volume of the multiparticle system. This results
immediately in the "sharp"{} energy-momentum conservation and
presence of the $\delta$-function $\delta^4(K - P)$ as a factor of
the $S$-matrix element (here, $K=k_1+k_2$ is the total momentum of
the pion pair, and $P=p_+ +p_-$ is the total momentum of the
lepton pair). As a consequence of the equality $K=P$, the
invariant mass of the two-pion system is specified as $K^2=M^2$,
where $M$ is the invariant mass of a registered lepton pair, i.e.
$P^2=M^2$. However, if the pion gas does not spread in all the
space, but it occupies a finite space-time volume, and its
lifetime is not long enough, the $\delta$-function should be
smeared to another distribution, e.g., $\rho(K - P)$ which
represents the distribution of the two-pion total energy $K^0$ and
total momentum ${\bf K}$ around the measurable quantities $P^0$
and ${\bf P}$, respectively. Effectively, this means the smearing
of the pion pair invariant mass $M_\pi=\sqrt{K^2}$ around the mass
$M$ of a lepton pair. For instance, in the case of a finite time
interval, one can use the Breit-Wigner energy distribution in
place of $\delta(K^0 - P^0)$.
In the case of a finite spatial volume, one can introduce a
relevant form-factor $\rho( {\bf K}- {\bf P})$ instead of $
\delta^3( {\bf K}- {\bf P} ) $. The consequence of smearing a
distribution in the 4-momentum space is strong enough. The rate
becomes measurable under the "standard" \, threshold, i.e. for the
values $M < 2m_\pi$, and down to two lepton masses. For instance,
for the electron-positron pair production, the rate becomes
measurable even in the vicinity of $2m_e\approx 1$~MeV, where
$m_e$ is the electron mass. Obviously, this phenomenon is due to
the uncertainty principle. In fact, if the life-time of a pion
system is restricted, for instance, by $\tau =2\, $fm/c, then the
energy (invariant mass) uncertainty is in the range up to $\Delta
E=100\, $MeV. Actually, this means that lepton pairs with
invariant masses down to $M \propto 2m_\pi-100$~MeV can be
registered. Indeed, in accordance with quantum mechanics, we can
speak about this matter because the experimental data are taken
from every particular event of the collision of two relativistic
nuclei. Reactions (quark-antiquark annihilation, pion-pion
annihilation, etc.) attributed to a particular event take place
during a finite time ($\tau=4\div 10~$fm/c) and in a bounded
volume ($R=4\div 10~$fm) which are fixed by a number of other
measurements characterizing the particular event.

Another important consequence of the finiteness of a reaction
space-time volume is a cutting of the "exact connection"{}
between the lepton-lepton c.m.s. and the pion-pion c.m.s. Indeed,
the relation between the total momentum $P$ of an outcoming lepton
pair and the total momentum $K$ of an incoming pion pair is
weighted now by the distribution function $\rho(K - P)$ which can
be regarded as a smeared pattern of $\delta^4(K - P)$. That is
why, any quantity determined (evaluated) in the lepton-lepton
c.m.s. which moves with the velocity $ {\bf v}_P = {\bf P}/P_0 $
in the lab system should be Lorentz-transformed to the pion-pion
c.m.s. which, in turn, moves with the velocity ${\bf v}_K= {\bf
K}/K_0$ in the lab system. As seen, such a transformation brings a
sufficient increase of the electron-positron emission rate for
small invariant masses $M$ which are in the range $2m_e < M <
2m_\pi$.

Actually, a restriction of the space-time volume in the process
$a\, \overline{a} \to b\, \overline{b}$ induces several effects which should be
taken into account.
Basically, the nonequilibrium behavior results in a correction of the
standard spectrum \cite{wong} obtained in the infinite space-time.
On the other hand, this correction can be
conventionally separated into three different contributions:
a)~First, it is a nonstationary behavior of a multiparticle system.
If, for instance, the multipion system which provides pions for
the process
$\pi^+\, \pi^- \to l\, \overline{l}$ has a finite lifetime, then individual
pion states are also nonstationary, and a decay of the states should be taken
into account,
b) The second contribution comes from the direct presence of the form-factor
of the multipion system which practically determines the finiteness of
a space-time volume occupied by the system,
c) At least, as we discussed above, the third contribution comes from
smearing  of
the $\delta$-function connection between the total 4-momentum of incoming
particles (pions) and the total 4-momentum of outgoing particles (leptons).

The goal of the present paper is to estimate the influence of the finite
space-time volume of a hadron reaction zone on the dilepton emission rate.

\section{Two-pion annihilation in a dense hadron medium}

We assume that the pion plasma which has been formed after the
equilibration process exists in a finite volume, and the
confinement of pions to this volume is a direct consequence of the
dense hadron environment which prevents the escape of pions during
some mean lifetime $\tau.$ We sketch a possible geometry in Fig.
1. A small circle of radius $R$ represents the subsystem of pions
which is in the local equilibrium and moves with collective
velocity $\mathbf{v}$. If the fireball size is of the order of
$R_0\propto 4\div 10$ fm, then the mean size of a small subsystem
is of the order of $R\propto 1\div 5$ fm. So, we assume that pion
states $\phi_\lambda(x)$ ($\lambda$ accumulates relevant quantum
numbers) which can take part in the annihilation reaction are
effectively bounded in a finite volume marked by radius $R$ in
Fig.~1.

\noindent
\begin{center}
\epsfxsize=0.5\columnwidth \epsffile{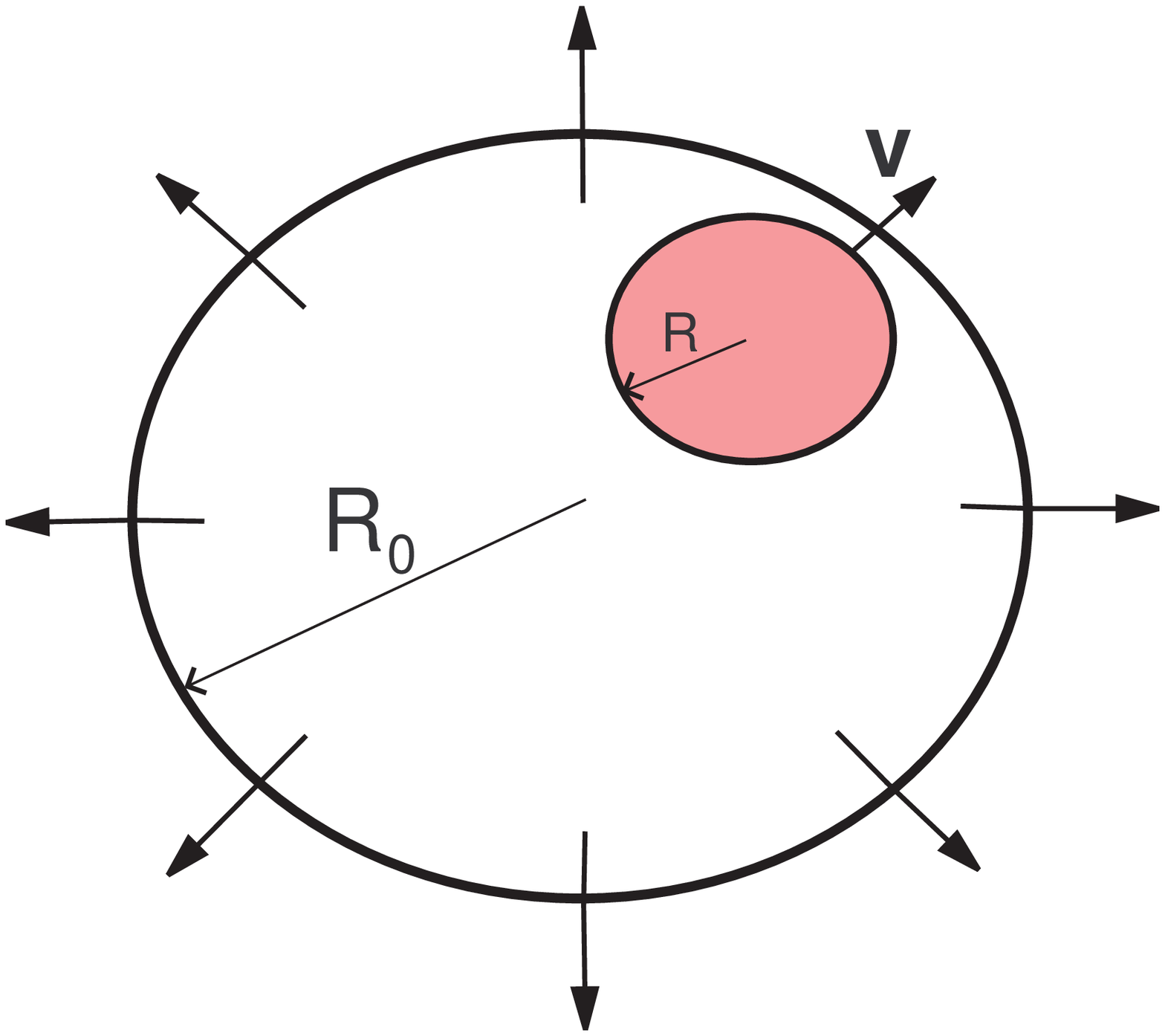}
\end{center}

\vspace*{-5pt}%

\noindent {\footnotesize Fig. 1. Sketch of an expanding fireball.
The small circle of radius $R$ represents the subsystem of pions
which is in a local equilibrium state and moves with the
collective velocity ${\bf v}$. If the fireball size is of the
order of $R_0 \propto 4\div 10$~fm, then the size of the small
subsystem is  $R \propto 1\div 5$~fm.

}
\vspace*{10pt}%

The pion-pion annihilation reaction in accordance with vector
meson dominance runs in the following way (see Fig.~2a, 2b): pions
annihilate through a $\rho $-meson which, in turn, transforms to a
virtual photon creating a lepton-antilepton pair, i.e. $ \pi^{+}\,
\pi^{-} \to \rho \to \gamma^{*} \to l\, \overline{l} $. Let us
point out that the system of pions under discussion has been
formed in ultrarelativistic nucleus-nucleus collisions and,
together with other species, represents a superhigh dense state of
matter, where the created pions are bounded during the lifetime of
the fireball by the hadron environment. To represent the finite
lifetime and spatial squeezing of pion states, which can take part
in the annihilation reaction, we adopt the ansatz for pion wave
functions
\begin{equation}
\phi_\lambda({\bf x})
=
\sqrt{\rho({\bf x})} e^{ i \alpha_\lambda({\bf x}) }
\, ,
\label{aa}
\end{equation}
where $\rho({\bf x})$ is the density of pions in the volume $V_\pi$,
in which pions being in a local thermodynamic equilibrium (marked by
radius $R$ in Fig. 1).
Assuming a single-particle wave function in the form (\ref{aa}),
we should like to note that particles which are squeezed
approximately
to a dense packing create a strongly correlated quantum system like
that observed in quantum liquids.
One of the recent results of HBT interferometry (the investigation of
two-particle spectra) indicates that the pions registered as
secondaries in heavy ion collisions
are emitted from the region of homogeneity of a fireball.
We take this as a basis for the assumption that the volume $V_\pi$
during the evolution inside a fireball
is approximately equal to that of the homogeneity region.
Hence, one can assume a smooth behavior of the phase of the wave
function (\ref{aa}) inside this volume and
expand the phase functions
$\alpha_\lambda(\bf{x})$ in the coordinate space at the center of the
region
\begin{eqnarray}
\alpha_\lambda(\bf{x})
&=&
\alpha_k
+
{\bf k}\cdot {\bf x} + \ldots,
\label{2-a}
\end{eqnarray}
where $\alpha_k \equiv \alpha_\lambda({\bf x}=0)$, ${\bf
k}={\displaystyle \partial \alpha_\lambda({\bf x}=0) }/ {
\displaystyle \partial  {\bf x} }$, and we start to label states
by the index $k$ in place of $\lambda$. For the sake of clearness
of the effects under investigation, we assume the free dispersion
law $\omega_k=\sqrt{m^2+{\bf k}^2}$. The quantities ${\bf k}$ and
$\omega_k$ can be interpreted as quasi-momentum and quasi-energy.
The amplitude $A_{\mathrm{fi}}\left( k_1,k_2;p_+,p_- \right) $ of
the annihilation process depicted in Fig.~2 reads
\begin{eqnarray}
&&\hspace{-0.65cm}
A_{\mathrm{fi}}\left( k_1,k_2;p_+,p_- \right)
=
\langle
{\bf p}_+,{\bf p}_- \left| S^{(2)} \right| {\bf k}_1,{\bf k}_2
\rangle
\nonumber \\
&&\hspace{-0.65cm}=
 -  \int d^4x_1\, d^4x_2 \,
\langle {\bf p}_+,{\bf p}_- \left|
  \,  T \left[  {\it H}_I^\pi(x_1) \, {\it H}_I^l(x_2)  \right] \,
      \right| {\bf k}_1,{\bf k}_2 \rangle
\nonumber \\
&&\hspace{-0.65cm}=
  i e^2 \int d^4x_1\,
\langle 0,~0 \left| j^\pi_\nu(x_1) \right| {\bf k}_1,{\bf k}_2 \rangle
\nonumber \\
&&\times \int d^4x_2 \,
\langle {\bf p}_+,{\bf p}_- \left| j^l_\mu(x_2) \right| 0,~0 \rangle
D^{\mu \nu }_F (x_1-x_2)
\, ,
\label{2-0}
\end{eqnarray}
where

$j^\pi_\mu(x)\! =\! -i \phi (x)
 \stackrel{\leftrightarrow}{\partial_\mu} \phi^+(x)$, \,
$j^l_\mu(x) \! = \! \overline{\psi } (x) \gamma_\mu \psi (x), $

\noindent and the photon propagator is
$$
D^{\mu \nu }_F (x_1-x_2) =
\int \frac{d^4P}{(2\pi )^4 }
\frac{g^{\mu \nu }}{P^2+i \epsilon }
e^{-i P\cdot (x_1-x_2) }
\ .
$$
To reflect explicitly the finiteness of the space-time volume of the
pion system, one can extract the pion 4-density $\rho (x_1)$ and write it
separately, so that the vertex $x_1$ in Fig.~2a, 2b is weighted by
this density.

\bigskip
\noindent \epsfxsize=1.1\columnwidth \epsffile{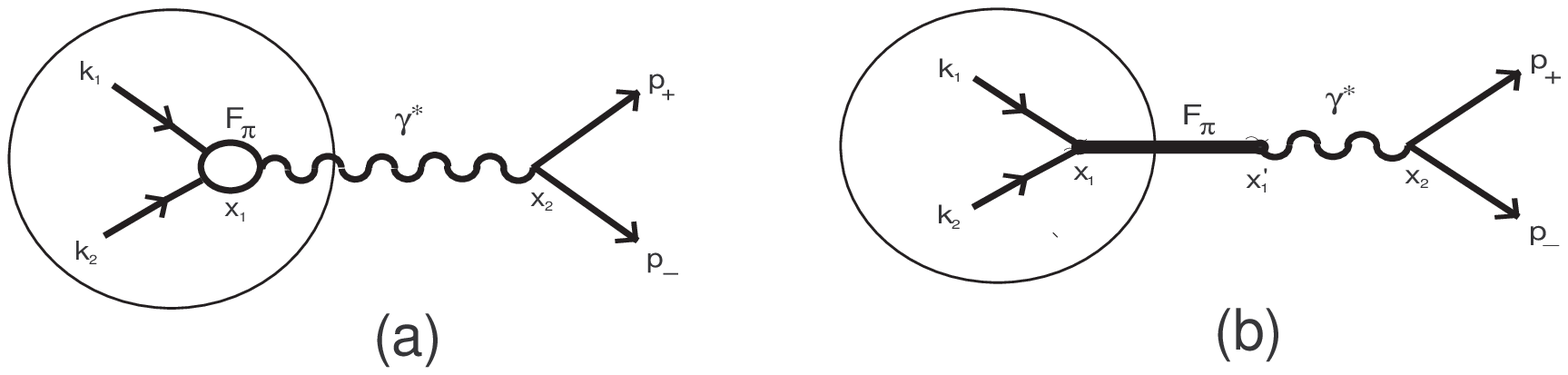}

\noindent {\footnotesize Fig. 2. The first-order nonvanishing
Feynman graph of the lepton pair creation  in the process $
\pi^{+}\, \pi^{-} \to \rho \to \gamma^{*} \to  l\, \overline{l} $.
The circle which surrounds the vertex $x_1$ sketches the finite
space-time region of the pion-pion interactions.

}
\vspace*{10pt}%

\noindent Then, the amplitude  can be rewritten in the form
\begin{eqnarray}
&&\hspace{-.5cm}A_{\mathrm{fi}}\left( k_1,k_2;p_+,p_- \right)
=
4\pi \alpha\! \int\!\! \rho(x_1) d^4 x_1 \! \int\!\! d^4 x_2 \,
\phi^{*}_{k_1}(x_1)
\nonumber \\
&&\hspace{-.5cm} \times  \phi ^{*}_{k_2}(x_1)\, D_{F}(x_1\!\!-x_2)
\,F_{\pi }\!\left( (k_1+k_2)^2 \right)
\phi _{p_+}(x_2)\, \phi _{p_-}(x_2) \,
\nonumber \\
&&\hspace{-.5cm} \times  m_{\mathrm{fi}}\left( k_1,
k_2;p_{+},p_{-} \right)\!\!,\qquad
\label{2-02}
\end{eqnarray}
where
$\alpha =e^2/ 4\pi$ and $\phi_k (x)=[(2\pi)^32E_k]^{-1/2}\exp{(-ik\cdot x)} $
with $k=k_1$, $k=k_2$ for pion states and with
$k=p_{+}$, $k=p_{-}$ for final lepton states (the lepton spinor part is
hidden in the matrix element $m_{\mathrm{fi}}$),
where $p^0_{\pm}=E_{\pm}=\sqrt{m_e^2+{\bf p}_{\pm}^2}$.
The quasi-energy of incoming pions is
defined as $k^0_i=E_i=\sqrt{m_\pi^2+{\bf k}_i^2}$ with
$i=1,2$. A $\rho $-meson form-factor $F_\pi (M)$ is taken in the form
\cite{gale87,gounaris}
\begin{equation}
\left| F_\pi(M^2) \right| ^2 =
\frac{ m_\rho^4 }{ (M^2-{m'}_\rho^2)^2
  +m_\rho^2 \, \Gamma_\rho^2 }
\label{2-03}
\end{equation}
with
$m_\rho=775\, {\rm MeV}, \ {m'}_\rho=761\, {\rm MeV}, \
\Gamma_\rho=118\, {\rm MeV}$.

\noindent The standard calculation of the matrix element
$m_{\mathrm{fi}}$ gives
\begin{eqnarray}
| m_{\mathrm{fi}}|^2
&=&
\sum_{\mathrm{spins}} t_\nu^+ t_\mu (k_1-k_2)^\nu (k_1-k_2)^\mu
\nonumber \\
&=&  8(k_1-k_2)\cdot p_{+}\ (k_1-k_2)\cdot p_-
\nonumber \\
&& - 4(k_1-k_2)^2(m_e^2+p_+\cdot p_-)
\ ,
\label{2-04}
\end{eqnarray}
where $t_\nu = \overline{u}(p_-)\gamma_\nu v(p_+)$ with $u(p)$ and $v(p)$
being the electron and positron spinors, respectively.
In the momentum space representation, amplitude (\ref{2-02}) looks like
\begin{eqnarray}
&&\hspace{-0.7cm}
A_{\rm fi}\!\left(k_1,k_2;p_+,p_- \right)
=
\frac{\displaystyle \alpha F_{\pi }\left( (k_1+k_2)^2 \right)}
     {\displaystyle 2\pi^2 \sqrt{2E_{1}2E_{2} }}
     \rho (k_1\!\!+k_2\!\!-\!\!p_+\!\!-\!\!p_-)
\nonumber\\
&&\hspace{2cm}
     \times \,D_F(p_+ + p_-) \frac{m_{\mathrm{fi}}(k_1,k_2;p_+,p_-)}
              {\displaystyle (2\pi )^{3}\sqrt{2E_+ 2E_- }},
\label{2-3}
\end{eqnarray}
where
$\rho (k)\! =\! \int d^{4}x \rho(x) e^{ik\cdot x} $
is the Fourier transform of the pion density.
The quantity $\rho (k_1\! +k_2\! -\! p_+\! -\! p_-)$ in
(\ref{2-3}) stands obviously in place of the delta function
$(2\pi)^4 \delta^4(k_1\! +k_2\! -\! p_+ \! - \! p_- )$ as a
reflection of the finiteness of the space-time volume occupied by
the pion system.

We consider two models (a) and (b) of the pion-pion annihilation,
and each of them can be put in correspondence to Fig.~2a or
Fig.~2b, respectively. Because we concentrate on the investigation
of the influence of the features of annihilating pion states on
the dilepton emission, we take the vacuum width and mass of the
$\rho$-meson form-factor in (\ref{2-03}) for both models. At the
same time, to reflect in some way the influence of the dense
hadron environment on the propagation of a $\rho$-meson, we adopt
in model (a) a zero mean free path of $\rho$-mesons what is
reflected in the diagram, see Fig.~2a. Amplitude (\ref{2-02}) was
written in accordance with model (a). On the other hand, in the
standard vector meson dominance treatment reflected by Fig.~2b,
the explicit propagation of a $\rho$-meson is taken into account.
The formal difference of these two approaches reveals itself just
as the difference of the arguments of the $\rho$-meson propagator
$F_\pi(k^2)$ which is $F_\pi\left( (k_1+k_2)^2 \right)$ and
$F_{\pi }\left( (p_++p_-)^2 \right)$ for models (a) and (b),
respectively. It is worth to note that models (a) and (b) coincide
when there is no restriction on the 4-volume of the reaction zone.
But it is not the case when the reaction takes place in a finite
space-time volume. Below, we analyze the consequences of the
difference of these two approaches for the rate of dilepton
emission.

The number of lepton pairs produced per event in the element of the
dilepton 4-momentum space $d^4P$  from the pion momentum space
elements $d^3k_1$ and $d^3k_2$ reads
\begin{eqnarray}
\frac{\displaystyle dN^{(\rho)} } {\displaystyle d^3k_1\, d^3k_2\,
d^4P} \, \, =  \! \! && \! \int \! d^3p_{+} \, d^3p_{-} \,
\delta^4(p_{+}+p_{-}-P)
\nonumber \\
&& \! \! \times \, \big| \, A_{\rm fi} \left( k_1,k_2;p_+,p_- \right) \big|^2
\, ,
\label{2-4}
\end{eqnarray}
where the super index $(\rho)$ denotes the presence of the pion source
form-factor $\rho(K-P)$.
Taking an explicit form of amplitude (\ref{2-3}), the expression on the
r.h.s. of (\ref{2-4}) can be represented as
\begin{equation}
\frac{ \displaystyle dN^{(\rho)} }
{ \displaystyle d^3 k_1 d^3 k_2 d^4P }
=
\left| \rho (k_1+k_2-P) \right| ^2 \,
\frac{\displaystyle dN}
{\displaystyle d^3 k_1 d^3 k_2 d^4P} \ ,
\label{2-5}
\end{equation}
where $dN/d^3k_1d^3k_2d^4P$ is the "standard"{} rate of particle
pair production with the total momentum $P$,
\begin{eqnarray}
&&\hspace{-0.5cm}\frac{\displaystyle dN} {\displaystyle d^3k_1\,
d^3k_2\, d^4P} = \frac{\displaystyle \alpha^2\, |F_\pi\left(
(k_1+k_2)^2 \right)|^2}
              {\displaystyle 4\pi^4\, 2E_1\, 2E_2\, P^4 }\nonumber\\
&&\hspace{-0.5cm}\times \int \frac{d^3p_{+}}{(2\pi )^3\, 2E_{+} }
     \frac{d^3p_{-}}{(2\pi )^3\, 2E_{-} } \left|
     m_{\mathrm{fi}}\right| ^{2} \delta^4(p_{+}+p_{-}-P).
\label{2-6}
\end{eqnarray}
Meanwhile, there are new important features relevant to
(\ref{2-6}) which are the consequence of the space-time
restrictions on the vertex $x_1$ in the diagrams depicted in
Fig.~2. Indeed, in the evaluation of the expression on the r.h.s.
of (\ref{2-6}), it should be taken into account the relative
movement of the pion-pion c.m.s. and the lepton pair c.m.s. For
instance, velocity of the pion-pion c.m.s. is ${\bf v}_K= {\bf
K}/K_0$, where $K=k_1+k_2$, whereas the dilepton c.m.s. moves with
velocity ${\bf v}_P= {\bf P}/P_0$. That is why, the velocities
${\bf v}_K$ and ${\bf v}_P$ are not identical now, as it would
have been in the presence of the delta-function $ \delta^4 (K-P) $
when the annihilation reaction takes place in the infinite
space-time. These two center-off-mass systems are "disconnected"{}
now and the relation between the total momenta $K$ and $P$ is
weighted by the form-factor $\left| \rho (K-P) \right| ^2$ which
stands as a prefactor on the r.h.s. of (\ref{2-5}).

Actually, the violation of the energy-momentum conservation is
typical of quantum physics, when the process or reaction under
consideration takes place during a finite time interval (energy
uncertainty) or in a finite coordinate space region (momentum
uncertainty). In essence, any system with finite lifetime reveals
a resonance-like behavior, and the energy-smearing function $\rho
(K^0-P^0)$ can look, hence, like the Breit-Wigner function
(Lorentz shape) $\rho(K^0~-~P^0)=\frac{1}{\pi}\,
\frac{\Gamma}{(K^0-P^0)^2+\Gamma^2}$ with $\Gamma=1/\tau$ as the
width of the energy probability distribution ($\tau$ is the
lifetime) or like the Gaussian function
$\rho(K^0~-~P^0)=(\tau/\sqrt{\pi})\exp{[-(K^0-P^0)^2 \tau^2]}$. In
these probability distributions, $P^0$ represents the mean value
as an external quantity, and $K^0$ is a running value of energy
which is distributed around $P^0$. In our case, the total energy
of the pion pair $K^0$ is distributed around a fixed value $P^0$
(experimentally measured dilepton total energy), and the smearing
is a consequence of the finiteness of the pion system lifetime.

Substituting the result of the evaluation of the integral on the
r.h.s. of (\ref{2-6}) in (\ref{2-5}), we obtain
\begin{eqnarray}
&&
\hspace{-0.65cm}
\frac{ \displaystyle dN^{(\rho)} }
     { \displaystyle d^3k_1\, d^3k_2\, d^4P } \,
= \,
  \left| \rho (k_1+k_2-P) \right|^2
  \frac{ \left( {\bf k}_1-{\bf k}_2 \right)_P^2 }{ 3(2\pi)^5 }
\nonumber \\
&&
\hspace{-0.65cm}
\times \frac{\displaystyle \alpha^2
       \left| F_\pi \left( (k_1+k_2)^2 \right) \right|^2 }
       {\displaystyle 4\pi^4\, 2E_1\, 2E_2\, P^2 }  \!
       \left( 1+\frac{2m_e^2}{P^2} \right) \! \!
       \left( 1-\frac{4m_e^2}{P^2} \right)^{1/2}
,
\label{2-7}
\end{eqnarray}
where integration was done in the lepton pair c.m.s.
Subindex $P$
denotes that the quantity $ ({\bf k}_1-{\bf k}_2)^2 $ is taken in
the P-system which is, by definition, the lepton pair c.m.s.,
whereas, subindex $K$ means that the quantity is taken in the
K-system which is the pion pair c.m.s.
As we have discussed above, such a separation is due to the fact that
the velocity ${\bf v}_K$ of the K-system and that ${\bf v}_P$ of the
P-system do not coincide.
For the further calculation, the quantity $ ({\bf k}_1-{\bf k}_2)_P^2 $
will be Lorentz-transformed to $ ({\bf k}_1-{\bf k}_2)_K^2 $.
For this purpose, we start from the covariant relation
$q_P^2 = q_K^2 $,
where the shorthand notation $q=k_1-k_2$ is adopted.
Since
$ \left( q_P^0 \right)^2 = (q\cdot P)^2/P^2 $,
one can write
\begin{equation}
{\bf q}_P^2 = {\bf q}_K^2 \left( 1+ \frac{{\bf P}_K^2}{P^2}
\cos^2{\theta} \right)
\, ,
\label{a15}
\end{equation}
where $\theta $ is the angle between vectors
${\bf q}$ and ${\bf P}$
when they are considered in the pion pair c.m.s. Taking into account
that
${\bf P}_K^2 = \left( P_K^0 \right)^2 - P^2 =
 \frac{(P\cdot K)^2}{K^2} -P^2 $,
and substituting this relation in (\ref{a15}), we get the relation
\begin{equation}
\left( {\bf k}_1 - {\bf k}_2 \right)_P^2
=
\left( {\bf k}_1 - {\bf k}_2 \right)_K^2 \!
\left[ 1\!+\! \cos^2{\theta}
       \left(\! \frac{(P\cdot K)^2}{P^2\, K^2}\! -\!1 \! \right)
       \right]
\!\! .
\label{a17}
\end{equation}
So, the expression on the r.h.s. of (\ref{a17}) is related to the
pion pair c.m.s.

The next step of evaluations needs the averaging of the quantity
$ dN^{(\rho)}/d^3k_1\, d^3k_2\, d^4P $ (\ref{2-7}) over the pion
momentum space with the distribution function which reflects the model
of a pion system created in a relativistic heavy ion collision.
For our estimation, we take the widely used thermodynamic model:
the subsystem of pions which is under consideration is in the thermal
equilibrium characterized by the temperature $T$.
Then, one has to weight the incoming pion momenta by a thermal
distribution function $f_{\rm th}(E)$ and make integration with
respect to these momenta:
\begin{eqnarray}
\left\langle \frac{ \displaystyle dN^{(\rho)} }{ \displaystyle d^4 P }
\right\rangle\!
=
\!\int\!\! d^3k_1 f_{\rm th}(E_1)\int d^3k_2~f_{\rm th}(E_2)
\nonumber \\
\times \frac{ \displaystyle dN^{(\rho)} }
            { \displaystyle d^3k_1d^3k_2d^4P }.
\label{2-8}
\end{eqnarray}
It is pedagogical to represent this equation in the form where the
explicit
factorization of the pion source form-factor $|\rho(K-P)|^2$ is made:
\begin{eqnarray}
\left \langle
\frac{ \displaystyle dN^{(\rho)} }{ \displaystyle d^4P }
\right \rangle
=
\int d^4 K~\left| \rho (K-P)\right|^2
\left \langle
\frac{\displaystyle dN}{\displaystyle d^4 K~d^4 P}
\right \rangle
\ .
\label{2-9}
\end{eqnarray}
Here, we introduce the auxiliary quantity
\begin{eqnarray}
\left \langle \frac{\displaystyle dN}{\displaystyle d^4K~d^4P}
\right \rangle \equiv \int d^3 k_1~f_{\rm th}(E_1)\int d^3
k_2~f_{\rm th}(E_2)\nonumber\\
\times \frac{\displaystyle
dN}{\displaystyle d^3 k_1 d^3 k_2 d^4P} \delta^4 (k_1+k_2-K) \, ,
\label{2-10}
\end{eqnarray}
which is the averaged dilepton number density with respect to the
total 4-momenta $K$ and $P$. In this formula, the spectrum $dN/d^3
k_1 d^3 k_2 d^4P$ equals the expression on the r.h.s. of
(\ref{2-7}) without the first factor $\left| \rho (k_1+k_2-P)
\right|^2$. The meaning of the averaging on the r.h.s. of
(\ref{2-10}) is quite transparent: we integrate over pion momenta
with weights which are the thermal distribution functions of pions
in the momentum space, keeping meanwhile the total energy-momentum
of the pion pair as a constant equal to $K$. The resulting
quantity on the l.h.s. of (\ref{2-10}) looks like we cut the
$s$-channel of the reaction $\pi^+\pi^- \to l\overline{l}$ into two
independent blocks (vertices) (see the diagrams depicted in
Fig.~2) and calculate the distribution $\langle dN/d^4 K
d^4P\rangle $ with respect to the incoming total momentum $K$ and
outgoing total momentum $P$ independently. Then, Eq. (\ref{2-9})
gives the connection of these vertices with a weight function, the
form-factor $ \left| \rho (K-P)\right|^2 $, by reflecting the fact
that the pion system lives in a finite space-time volume. On the
other hand, one can regard Eq. (\ref{2-9}) as an averaging of the
{\it random} quantity $\left<\frac{dN}{d^4Kd^4P}\right>$  with the
help of the distribution function $|\rho(K-P)|^2$, which is
centered around the mean value $P$. In this sense, the hadron
medium temporary holding pions in a local spatial region (this
generates the local pion distribution $\rho(x)$) randomizes the
system of pions on the quantum level. This randomization of the
pion pair total momentum $K$ is really a quantum one as a
particular manifestation of the uncertainty principle. Note that
the thermal randomization of the multipion system was included
already to the quantity $\left< \frac{dN}{d^4Kd^4P}\right>$  (see
(\ref{2-10})) through the thermal distribution functions
$f_\mathrm{th}(E)$.

Taking into account a result of integration of the expression on
the r.h.s. of (\ref{2-10}),
we can rewrite (\ref{2-9}) as
\begin{eqnarray}
&&\hspace{-.5cm}
\left\langle \frac{ \displaystyle dN^{(\rho)} }{\displaystyle d^4P }
\right\rangle
=
  \frac{ \displaystyle \alpha^2 C^2 }{ \displaystyle 3(2\pi )^8 }
  \left( 1- \frac{4m_e^2}{P^2} \right)^{1/2}
  \left( 1+ \frac{2m_e^2}{P^2} \right)\,
\nonumber
\\ &&
  \times \int d^4 K \left| \rho (K-P) \right|^2 e^{-\beta u\cdot K}
  \frac{K^2}{P^2}\, \big| F_\pi(K^2) \big|^2
\nonumber \\
&&
  \times \left( 1- \frac{4m_\pi^2}{K^2} \right)^{3/2}
  \left[
       1 + \frac{1}{3}
       \left( \frac{ \left(P\cdot K\right)^2 }{P^2\, K^2} -1\right)
  \right]
\, ,
\label{2-12}
\end{eqnarray}
where $u$ is the collective 4-velocity of the pion system, which is an
element of the fireball.
We take the Boltzmann distribution
$f_{\rm th}(E)=C\exp (-\beta E)$ as a pion thermal distribution
function with inverse temperature $\beta =1/T$, where $C$ is a
normalization constant.
The factor $1/3$ in the square brackets on the r.h.s. of (\ref{2-12})
appears after the integration over angles.
Expression (\ref{2-12}) is a final result for the distribution
of the number of particles as a function of dilepton 4-momentum $P$.
The further evaluation can be made for a particular form
of the pion system form-factor $|\rho (K-P)|^2$.

We add for completeness that, for the process which corresponds to
the diagram depicted in Fig.~2b [model (b)], one should correct the
last expression (\ref{2-12}) by shifting the pion form-factor which
is now taken in the form
$\big| F_\pi(P^2) \big|^2$, in front of the integral.

The convergence of (\ref{2-12}) to the standard result \cite{wong}
can be recognized, when one changes the pion system form-factor to
the $\delta$-function, namely $\rho (K-P) \to (2\pi)^4
\delta(K-P)$. It is easy to see that the expression in square
brackets can be immediately transformed to unity when $K=P$,
because there is no difference between the P- and K-systems in
this case. They both move now with the velocity ${\bf v}_K={\bf
v}_P={\bf P}/P_0$. Then, dividing by the 4-volume $\widetilde{T}V$
($\widetilde{T}$ is the time interval and $V$ is the volume), we obtain
the rate ($R\equiv N/\widetilde{T}V$)
\begin{eqnarray}
\hspace{-1cm} && \left\langle
     \frac{ \displaystyle dR }{ \displaystyle d^4P }
                 \right\rangle_{V\to \infty \atop \widetilde{T} \to \infty}
=
\frac{ \displaystyle \alpha^2 C^2 }{ \displaystyle
3(2\pi )^8 }\, e^{-\beta P_0} \big| F_\pi(P^2) \big|^2 \nonumber\\
&& \times
  \left( 1- \frac{4m_\pi^2}{P^2} \right)^{3/2}
  \left( 1- \frac{4m_e^2}{P^2} \right)^{1/2}
  \left( 1+ \frac{2m_e^2}{P^2} \right)
  .
\label{2-12a}
\end{eqnarray}
To obtain the distribution of the number of lepton pairs with respect
to the invariant mass $M$ of a pair, it is necessary to integrate distribution
(\ref{2-12}),
$ \left\langle dN^{(\rho)}/d^4P \right\rangle $,
over the 4-momentum space $P$ setting the value $P^2$ to $M^2$, namely
\begin{eqnarray}
\left \langle \frac{ \displaystyle dN^{(\rho)} }{ \displaystyle
dM^2 } \right \rangle = \int d^4 P \, \left \langle \frac{
\displaystyle dN^{(\rho)} }{ \displaystyle d^4P } \right \rangle
\, \delta ( P^2 - M^2 )\, \theta (P_0) \, . \label{2-14}
\end{eqnarray}
Hence, using our result (\ref{2-12}), we can write the distribution of
the number of created lepton pairs $N^{(\rho)}$ with respect to the invariant
mass $M$ of a lepton pair as
\begin{eqnarray}
&&
\hspace{-0.6cm}
\left\langle \frac{ \displaystyle dN^{(\rho)} }
    {\displaystyle dM^2 } \right\rangle  \!
= \!
\frac{ \displaystyle \alpha^2 C^2 }{ \displaystyle 3(2\pi )^8 }
\left( 1- \frac{4m_e^2}{M^2}\right)^{\! 1/2} \!  \! \!
\left( 1+ \frac{2m_e^2}{M^2} \right)  \!
\int \! \! \frac{d^3 P}{2\, \Omega({\bf P}) }
\nonumber \\
&&
\hspace{-0.6cm}
\times
\int d^4K \;\theta\left(K_0\right) \,
\theta\left(K^2-4m_\pi^2\right) \left| \rho (K-P) \right|^2
e^{-\beta K_0} \frac{K^2}{M^2}
\nonumber \\
&&
\hspace{-0.6cm}
\times
\big| F_\pi(K^2) \big|^2 \! \! \left( 1-\frac{4m_\pi^2}{K^2} \right)^{\! 3/2} \! \!
\left[ 1 + \frac{1}{3} \left( \frac{\left(P\cdot K\right)^2 }
  {M^2\, K^2} -1 \right) \! \! \right]
\label{2-15a}
\end{eqnarray}
with $P_0=\Omega({\bf P}) \equiv \sqrt{M^2+{\bf P}^2}$. As for the
presence of the $\theta$-function
$\theta\left(K^2-4m_\pi^2\right)$ on the r.h.s. of (\ref{2-15a}),
we would like to stress that the integration with respect to $K$
is going on over the domain where the invariant mass of a pion
pair $M_\pi \equiv \sqrt{K^2}$ is not lesser than two pion masses.
On the other hand, possible finite values of the distribution $
\left\langle dN^{(\rho)}/dM^2 \right\rangle $ below the two pion
mass threshold can occur due to the smearing of the standard rate
which is generated by the quantum fluctuations of the pion pair
total momentum $K$ as a realization of the uncertainty principle.
These quantum fluctuations are visualized in (\ref{2-15a}) by the
presence of the pion system form-factor $|\rho (K-P)|^2$. Note, if
one follows model (b) which corresponds to the process depicted in
Fig.~2b, then the $\rho$-meson form-factor on the r.h.s. of
(\ref{2-15a}) should be taken in the form $ \big| F_\pi(M^2)
\big|^2 $ and shifted to the front of the integral. Equation
(\ref{2-15a}) is the main theoretical result of the present paper.
The next steps in the investigations of the rate (\ref{2-15a}) can
be made by taking a particular source form-factor $|\rho (K-P)|^2$
of the many-particle (multipion) system.

\subsection{ Gaussian pion system form-factor }

To make an access to the effects under investigation transparent as
much as possible, we take, as a model of the pion source, the Gaussian
distribution in space and the Gaussian decay of the multipion system
\begin{eqnarray}
\rho (x)
=
\exp{ \left( - \frac{t^2}{2\tau^2} - \frac{{\bf r}^2}{2R^2}  \right) }
\, ,
\label{2-14a}
\end{eqnarray}
which transforms to $1$ for the large enough mean lifetime $\tau$ and
mean radius $R$ of the system, i.e., when
$\lim_{\tau \to \infty \atop R \to \infty} \rho (x) = 1 $, amplitude
(\ref{2-02}) comes to the standard form.
The Fourier transformation of $ \rho (x) $ reads
\begin{equation}
\rho (Q)
=
 (2\pi)^2 \tau R^3 \exp{\left[-\frac{1}{2} \left(Q_0^2\tau^2
 +
 {\bf Q}^2R^2 \right) \right] }
\, .
\label{2-15}
\end{equation}
The correspondence with the form-factor in case of the infinite
space-time volume  should be installed as
\begin{equation}
\lim _{\tau,\, R \to \infty }\rho^2 (Q) =\widetilde{T} V~(2\pi )^4\delta^4 (Q)
\, ,
\label{2-16}
\end{equation}
where, on the r.h.s., one $\delta$-function is transformed to the 4-volume,
$\widetilde{T}$ is a time interval, and $V$ is a spatial volume.
On the other hand, for the l.h.s. of (\ref{2-16}), we have explicitly
\begin{eqnarray}
\lim _{\tau,\, R \to \infty }\rho^2 (Q) = (\pi^2 \tau R^3)
(2\pi)^4 \, \delta^4(Q) \, . \label{2-16a}
\end{eqnarray}
Hence, by comparing the right-hand sides of (\ref{2-16})
(\ref{2-16a}), we can conclude that the quantity $\pi^2 \tau R^3 $
should represent the 4-volume $\widetilde{T}V$ for the source
parametrization (\ref{2-14a}).

With the use of the above source distribution function,
Eq.~(\ref{2-15a}) gives the spectrum with respect to the lepton pair
invariant mass
during a unit time interval and from the unit volume of the multipion
system
\begin{eqnarray}
&& \hspace{-.7cm} \left \langle \frac{\displaystyle dN^{(\rho)} }{
\displaystyle \widetilde{T} V dM^2 } \right \rangle = \frac{
\displaystyle \alpha^2 C^2 }{ \displaystyle 3(2\pi )^4 }
\left( 1- \frac{4m_e^2}{M^2} \right)^{1/2} 
\left( 1+\frac{2m_e^2}{M^2} \right)
\nonumber \\
 &&\hspace{-.5cm}
 \times \, \int d^4 K \, \theta\left(K_0\right) \,
 \theta\left(K^2-4m_\pi^2\right) \, e^{-\beta K_0}
 \frac{K^2}{M^2}\, \big| F_\pi(K^2) \big|^2 \,
\nonumber \\
&& \hspace{-.5cm}
\times \,
\left( 1- \frac{4m_\pi^2}{K^2} \right)^{3/2}
\int \frac{d^3 P}{ 2\sqrt{M^2+{\bf P}^2} } \, \frac{\tau}{\pi^{1/2}}
e^{-\left( K_0-P_0 \right)^2 \tau^2}
\nonumber \\
&&\hspace{-.5cm}
\times \,
\frac{R^3}{\pi^{3/2}} \, e^{-\left( {\bf K}-{\bf P} \right)^2 R^2}
\left[ 1 + \frac{1}{3} \left( \frac{ \left(P\cdot K\right)^2}
       {M^2\, K^2} -1 \right)
\right] \, , \label{2-17}
\end{eqnarray}
where
$P_0=\Omega({\bf P}) \equiv \sqrt{M^2+{\bf P}^2} $ and
$\widetilde{T}V=\pi^2 \tau R^3 $.
We write the source form-factor exponents together with relevant
coefficients in
the form which shows the evident transformation to the $\delta$-functions
$ \delta\left( K_0-P_0 \right) $
and
$ \delta^3\left( {\bf K}-{\bf P} \right) $
when $\tau \to \infty$ and $ R \to \infty $, respectively.

After the integration over angle variables in both integrals on
the r.h.s. of Eq.~(\ref{2-17}), we obtain the rate ($R^{\rm
(\rho)}\equiv N^{(\rho)}/\,\widetilde{T}V$) of dilepton production in
the reaction $\pi^+\, \pi^- \rightarrow l\, \overline{l}$
\begin{eqnarray}
&&\hspace{-.7cm}
\left \langle
\frac{ \displaystyle dR^{\rm (\rho)}}{ \displaystyle dM^2 }
\right \rangle
=
\frac{\displaystyle \alpha ^{2}C^{2} }{\displaystyle 3(2\pi )^3 }
\left( 1- \frac{4m_e^2}{M^2} \right)^{\frac12} \! \!
\left( 1+ \frac{2m_e^2}{M^2} \right)\! \! \int_{-\infty }^{\infty } \!
       \! \frac{\overline{P} d\overline{P} }{P_0}
\nonumber \\
&&\hspace{-.7cm}
\times\!  \int_0^{\infty } \! \! \overline{K} d\overline{K}~\frac{R}{\pi^{1/2}}
e^{-(\overline{K} -\overline{P} )^2 R^2 } \! \!
\int _{ \sqrt{4m_{\pi}^2+\overline{K}^2}}^{\infty } dK_0 \; e^{-\beta K_0}
\nonumber \\
&&\hspace{-.7cm}
\times  \frac{\tau }{\pi^{1/2}} e^{-(K_0 - P_0)^2\tau^2 }
\frac{K^2}{M^2}\, \big| F_\pi(K^2) \big|^2
\,
\left( 1- \frac{4m_\pi^2}{K^2} \right)^{\frac32}\,
\nonumber \\
&&\hspace{-.7cm}
\times \!
\left[
     1\!+\!\frac{1}{3M^2 K^2} \! \left(\! \left( \! P\cdot K \right)^2
     \! - M^2 K^2 \! + \! \frac{ P\cdot K }{ R^2 } \! + \!
     \frac{1}{ 2 R^4 } \! \right) \!
\right]
\, ,
\label{2-18}
\end{eqnarray}
where we use the notations
$$
\hspace{-.7cm}
\overline{K} \equiv |{\bf K}| \, ,\ \ \overline{P} \equiv |{\bf P}| \, , \ \
P_0 = \Omega( \overline{P} ) = \sqrt{M^2+\overline{P}^2 } \, , \ \
$$

$$
\hspace{-.7cm}
K^2 = K_0^2-\overline{K}^2 \, , \ \
P\cdot K = \Omega( \overline{P} ) K_0 - \overline{P} \overline{K}
\, .
$$
Let us recall that the expression in square brackets on the r.h.s. of
(\ref{2-18}) is
a relativistic correction which arises when we perform the Lorentz
transformation of the quantity
 $({\bf k}_1-{\bf k}_2)^2$ from the lepton pair c.m.s. to
the pion pair c.m.s.
The influence of the finite spatial size of the system on the
correction discussed is explicitly represented by two terms
$\propto R^{-2}$ and $\propto R^{-4}$ which reflect the fact that the
relativistic correction is larger for a smaller pion system.
We note again that, within model (b) corresponding to the process
depicted in Fig.~2b, just one correction should be done in expressions
(\ref{2-17}) and (\ref{2-18}).
Formally, it is necessary to change the pion form-factor argument from
$K^2$ to $M^2$, and then it can be written in front of the integral
in the form
$\big| F_\pi(M^2) \big|^2$.

\subsection{Passage to the limits $R\to \infty$ and $\tau \to \infty$}

First, we consider the limit $R \rightarrow \infty $, which means that
the multipion system occupies the infinite volume in the coordinate
space, but lives for a finite lifetime in this state.
By performing the integration in Eq.~(\ref{2-18}) with the use of
the $\delta$-function $\delta(\overline{K}-\overline{P}) $, we obtain
\begin{eqnarray}
&&\hspace{-.7cm}
\left\langle
\frac{ \displaystyle dR^{\rm (\rho)}}{ \displaystyle dM^2 }
                                       \right\rangle_{R\to \infty}
=
\frac{\displaystyle \alpha^2 C^2 \tau }{\displaystyle 24 \pi^{7/2} }
\left( 1- \frac{4m_e^2}{M^2} \right)^{1/2} \! \!
\left( 1+ \frac{2m_e^2}{M^2} \right)
\nonumber \\
&&\hspace{-.7cm}
\times
   \int _0^{\infty } \frac{\overline{P}^2 d\overline{P}}
       {\Omega(\overline{P})}
   \int_{\sqrt{4m_{\pi }^2+\overline{P}^2} }^{\infty }dK_0
       e^{ -[K_0 - \Omega(\overline{P})]^2\tau^2 } e^{-\beta K_0}\frac{K^2}
       {M^2}
\nonumber \\
&&\hspace{-.7cm}
\times
\big| F_\pi(K^2) \big|^2 \! \left( 1 \! - \! \frac{4m_\pi^2}{K^2}
                            \right)^{3/2} \!
 \left[
  1+\frac{\overline{P}^2 \left( K_0\! -\! \Omega(\overline{P}) \right)^2}{3M^2 K^2}
 \right] \! \!
,
\label{2-20}
\end{eqnarray}
where $\Omega( \overline{P} )=\sqrt{M^2+\overline{P}^2 } \, , \
K^2=K_0^2-\overline{P}^2 $. It is reasonable to note that, due to a
finite lifetime $\tau$ of the multipion system and, thus, a finite
lifetime of annihilating pion states, the off-shell behavior of
the two-pion system is seen pretty well. Indeed, it is reflected
by the presence of the distribution $(\tau/\sqrt{\pi}) e^{ -[K_0 -
\Omega(\overline{P})]^2\tau^2 }$ of the fluctuating total energy $K_0$
of the pion pair around the nominal quantity
$\Omega(\overline{P})=\sqrt{ M^2+{\bf P}^2 }$ in the integrand on the
r.h.s. of (\ref{2-20}). At the same time, the discrepancy of $K_0$
and $P_0$ causes the difference of velocity $\mathbf{P}/K_0$ of
the $K$-system and velocity $\mathbf{P}/P_0$ of the $P$-system.
Hence, it determines the relativistic correction which is
manifested by the presence of square brackets on the r.h.s. of
(\ref{2-20}). Let us consider now another limit. We take the
infinite lifetime of the pion system ($\tau \! \rightarrow \!
\infty $), but a finite volume of the reaction region. This leads
to the equality of the total energies of lepton and pion pairs.
The result of integration over $K_0$ on the r.h.s. of (\ref{2-18})
with the help of the $\delta$-function $\delta
\left(K_0-\Omega(\overline{P}) \right)$ is not zero when the following
inequality holds:
\begin{eqnarray}
\sqrt{4m_{\pi }^2+\overline{K}^2} \leqslant \Omega\left( \overline{P}
\right) \ . \label{2-23}
\end{eqnarray}
If the measured dilepton invariant mass $M$ is larger than
$2m_\pi$, inequality (\ref{2-23}) results in a restriction from
above of the integration over $\overline{K}$. Then the integration is
carried out in the limits $[0,\overline{K}_{\rm max}]$, where
$\overline{K}_{\rm max} = \sqrt{ \overline{P}^2+ M^2 - 4m_\pi^2 }$. On the
other hand, if the measured dilepton invariant mass $M \leqslant
2m_\pi$, it is necessary to introduce a restriction on the limits
of integration over the total momentum of the lepton pair
$\overline{P}$. By keeping all this together, we write the rate of the
reaction $\pi^+\, \pi^- \rightarrow l\, \overline{l}$ in the case
of a finite spatial volume of the reaction region as
\begin{eqnarray}
&&
\hspace{-.7cm}
\left \langle \frac{ \displaystyle dR^{\rm (\rho)}}
   { \displaystyle dM^2 } \right \rangle_{\tau \to \infty}
= \int^\infty _{-\infty} d\overline{P} \, J\left(M,R,\overline{P} \right)
\nonumber \\
&& \hspace{.7cm} - \ \theta \left( 2m_\pi-M \right)
\int_{-\overline{P}_{\rm min} }^{ \overline{P}_{\rm min} } d\overline{P}
J\left(M,R,\overline{P}\right) \, , \label{2-24}
\end{eqnarray}
where
$\overline{P}_{\rm min} = \sqrt{4m_\pi^2 - M^2 }$ and
\begin{eqnarray}
&&\hspace{-.7cm}
J\left(M,R,\overline{P} \right)
=
\frac{\displaystyle \alpha^2 C^2 R}{\displaystyle 24(\pi )^{7/2} } \
\frac{ \overline{P}\, e^{-\beta \Omega(\overline{P})} }
     {\Omega\left( \overline{P} \right) }
\int_0^{\overline{K}_{\rm max}} \overline{K} d\overline{K}
\nonumber \\
&&\hspace{-.7cm}
\times \, e^{ -(\overline{K}-\overline{P} )^2 R^2 } \,
\frac{\displaystyle K^2 }{\displaystyle M^2}\,
\left|F_{\pi }(K^2)\right|^2
\left(  1- \frac{4m_\pi^2}{K^2} \right) ^{3/2}
\nonumber \\
&&\hspace{-.7cm}
\times
\left[
     \! 1\! + \frac{1}{3M^2 K^2} \!
        \left(\! \Omega^2\left( \overline{P} \right)
              \left( \overline{P}-\overline{K} \right)^2\! + \!
              \frac{ P\cdot K }{ R^2 }\! + \!
              \frac{1}{ 2 R^4 }\!
        \right)
\right]
\label{2-25}
\end{eqnarray}
with
$ \Omega( \overline{P} ) = \sqrt{M^2+\overline{P}^2 } \, , \ \ K^2 =
\Omega^2\left( \overline{P} \right)-\overline{K}^2 \, , \ \ P\cdot K =
\Omega^2\!\left(\! \overline{P} \right) - \overline{P} \overline{K}\ {\rm and} \
\overline{K}_{\rm max} = \sqrt{ \overline{P}^2\!+\! M^2\!-\! 4m_\pi^2 }\, .
$

For completeness, let us consider the rate when we take both limits,
$R\to \infty$ and $\tau\to \infty$.
Equation (\ref{2-20}) represents the rate in the case of infinite
spatial volume.
If, in addition, we consider a big enough lifetime of the multipion
system, i.e. if we take the limit $\tau\to \infty$,
this results in the appearance of the $\delta$-function
$\delta\left(K_0 - \Omega(\overline{K})\right) $ on the r.h.s. of
(\ref{2-20}), and the integration gives
\begin{eqnarray}
&&\hspace{-.7cm}
\left \langle
      \frac{ \displaystyle dR^{\rm (\rho)}}{ \displaystyle dM^2 }
\right \rangle_{R\to \infty \atop \tau\to \infty}
=
\frac{\displaystyle \alpha^2 C^2 }{\displaystyle 3(2\pi )^3 }
\left( 1- \frac{4m_e^2}{M^2} \right)^{1/2} \! \! \left( 1+
\frac{2m_e^2}{M^2} \right)
\nonumber \\
&&\hspace{.5cm}
\times
\big| F_\pi(M^2) \big|^2 \,
\left(
      1 - \frac{4m_\pi^2}{M^2}
\right)^{3/2}\, \frac{M}{\beta} \, K_1(\beta M)
\, ,
\label{2-22}
\end{eqnarray}
where $K_1(z)$ is a McDonald function of the first order.
So, we come to the standard result \cite{wong} for the rate
of dilepton emission in $\pi^+ \, \pi^-$ annihilation in the case of the
Boltzmann distribution of pions in the stationary infinite
multipion system.

\subsection{ Evaluations of the dilepton emission rate }

Before going to numerically evaluate the rate of dilepton emission,
let us briefly consider the geometry of a multipion system.

After the nuclei collision, the highly excited nuclear matter goes
through several stages. At the same time, all steps of the
particle-particle transformations take place against the
background of the spatial dynamics of the system. It is believed
that the total system of pions which were created after
hadronization expands in all directions, and it is a hard task to
find even quasi-equilibrated pions in the fireball taken as a
whole.

Let us, however, consider a locally equilibrium system of pions
which is a small part of the total system (fireball) and moves
with a collective velocity ${\bf v}$. We draw it schematically in
Fig.~1, where the arrows mean the radially directed velocities of
the expanding hadron matter. Just to be transparent in the
consideration of finite space-time effects, we adopt a
quasi-static picture. If, for heavy colliding nuclei, the fireball
is of the mean radius $R_0 \propto 4\div 10$~fm, what is known
from interferometry, then the spatial region occupied by pions
which are in local equilibrium is of the size $R \propto 1\div
5$~fm. This small region is depicted in Fig.~1 as a circle of
radius $R$ on the body of the fireball. Hence, the consideration
of the previous paragraphs can be attributed to the multipion
subsystem which is in the local equilibrium in a small spatial
region (on the scale of the fireball size) characterized by radius
$R$. The evolution of this multipion subsystem in time is
restricted by the mean lifetime $\tau$ which is not bigger than
the total lifetime of the fireball. We evaluate numerically
integrals (\ref{2-18}) for different sets of parameters $R$ and
$\tau$. Fig.~3 shows the creation rate $dR^{\rm
(\rho)}_{e^+e^-}/dM$ for electron-positron pairs appearing from a
hot multipion subsystem characterized by the temperature
$T=180$~MeV.

\vspace*{2pt}%

\begin{center}
\includegraphics[width=0.8\columnwidth,height=0.6\columnwidth]{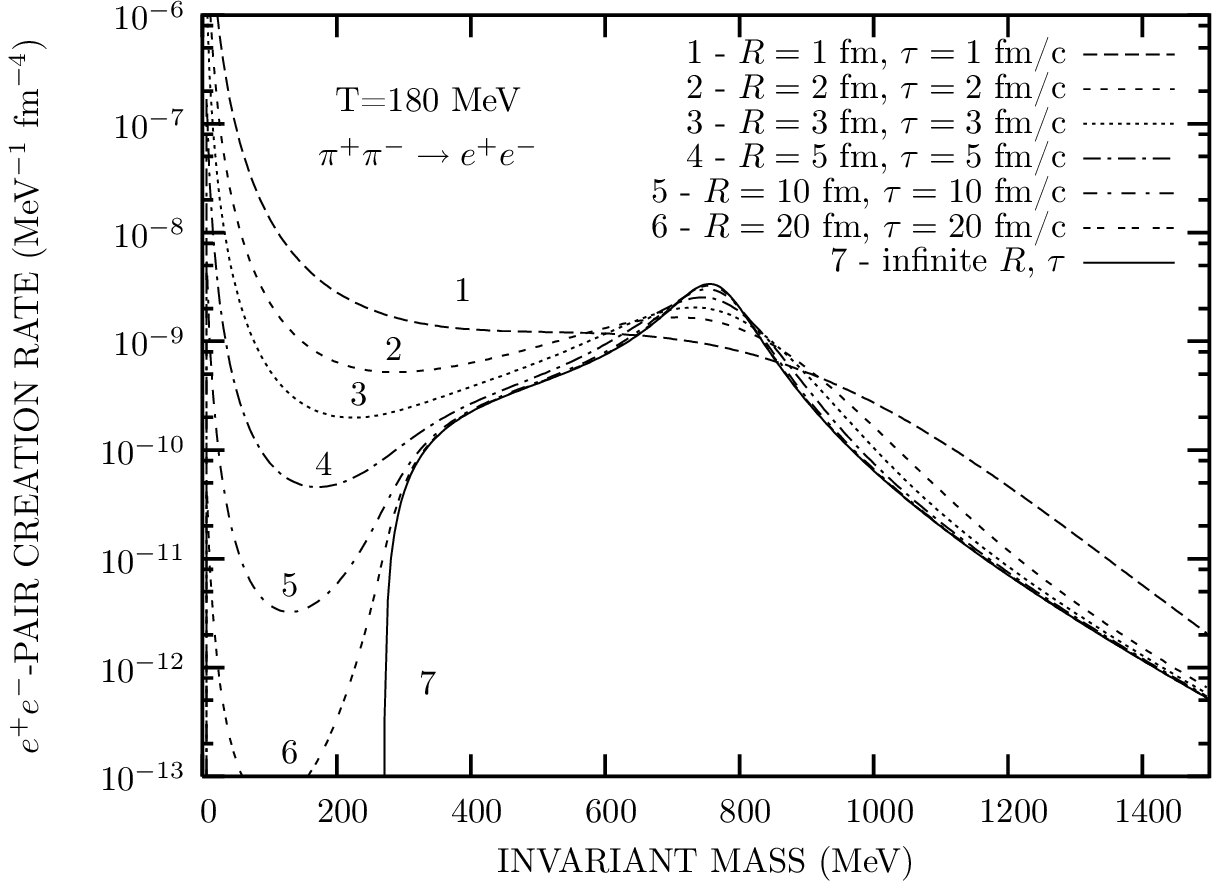}
\end{center}

\noindent {\footnotesize Fig. 3. The rate $dR^{\rm
(\rho)}_{e^+e^-}/dM$ of electron-positron pair creation in
pion-pion annihilation. Different curves correspond to the
different spatial sizes $R$ and different lifetimes $\tau$ of a
hot pion system, $T=180$~MeV. }
\medskip

\noindent There are seven curves in this figure which are labelled
by numbers $1, \ldots ,7$ in the following correspondence to the
values of the mean size of the pion subsystem and its lifetime:
$1)$ $R=1$~fm, $\tau=1$~fm/c; $2)$ $R=2$~fm, $\tau=2$~fm/c; $3)$
$R=3$~fm, $\tau=3$~fm/c; $4)$ $R=5$~fm, $\tau=5$~fm/c; $5)$
$R=10$~fm, $\tau=10$~fm/c; $6)$ $R=20$~fm, $\tau=20$~fm/c; $7)$
$R=\infty$, $\tau=\infty$. The evaluation for the infinite
space-time volume (see (\ref{2-22})), corresponds to curve $7$.
Notice how the $e^+e^-$ emission rate deviates from the standard
one (curve 7): it is finite in the region of the lepton invariant
mass $M\leqslant 2m_\pi$. As we discussed above, it is a
consequence of the quantum fluctuations of the total pion pair
momentum $K$ around the total lepton pair momentum $P$
(uncertainty principle). We emphasize as well that the tendency of
curves 1, 2 in Fig.~3 is similar to CERES data
\cite{agakichev95,agakichev98}. In Fig.~4, we show the evaluation
of the creation rate $dR^{\rm (\rho)}_{\mu^+\mu^-}/dM$ for
$\mu^+\mu^-$-pairs at the same temperature $T=180$~MeV, as for
electron-positron creation. There are also seven curves in this
figure which are labelled by numbers from $1$ to $7$ in the same
correspondence to the values of the mean size $R$ and mean
lifetime $\tau$ of the pion subsystem as we take for the previous
figure. Notice that the stronger the deviation of the $\mu^+\mu^-$
creation rate for a finite pion system from that for an infinite
pion system, the lesser the spatial size and lifetime of the pion
system similarly to the case of $e^+e^-$ creation. Of course, it
is a reflection of the uncertainty principle which is formally
expressed by the presence of the distribution (pion source
form-factor) $|\rho(K-P)|^2$ as a factor of the integrand on the
r.h.s. of (\ref{2-9}). The presence of the form-factor of the pion
subsystem results in a broadening of the rate: the effect is
bigger when the parameters $R$ and $\tau$ are smaller. Evidently,
the biggest broadening was obtained for the pion subsystem with
mean radius $R=1$~fm and lifetime $\tau=1$~fm/c. In contrast to
the standard result, Fig.~4 shows that the threshold of
$\mu^+\mu^-$ creation is at the value of invariant mass $M_{\rm
thresh}^{\mu^+\mu^-}=2m_\mu \approx 211.3$~MeV.

\bigskip
\begin{center}
\includegraphics[width=0.8\columnwidth,height=0.6\columnwidth]{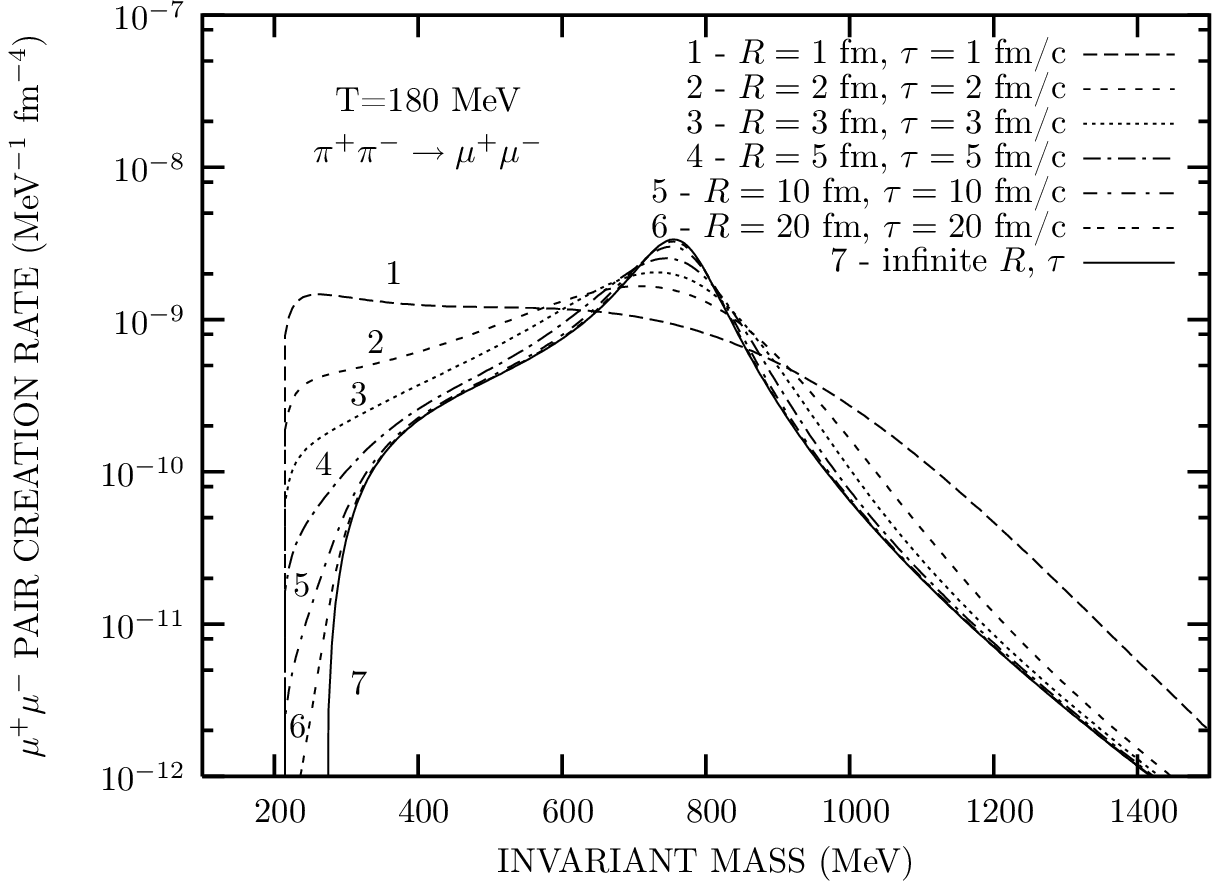}
\end{center}

\noindent {\footnotesize Fig. 4. The rate $dR^{\rm
(\rho)}_{\mu^+\mu^-}/dM$ of $\mu^+\mu^-$ pair creation in
pion-pion annihilation. Different curves correspond to the
different spatial sizes $R$ and different lifetimes $\tau$ of a
hot pion system, $T=180$~MeV. }
\vspace*{10pt}%

An analogous threshold for electron-positron creation, $M_{\rm
thresh}^{e^+e^-}=2m_e \approx 1.02$~MeV, is not been visible on
the scale of mass span which is taken in Fig.~3. In Fig.~5, we
show the creation rate $dR^{\rm (\rho)}_{e^+e^-}/dM$ for
electron-positron pairs in the region of the threshold for the
mean radius $R=1$~fm and lifetime $\tau=1$~fm/c of the pion
subsystem. The solid curve is a result of the evaluation of
expression (\ref{2-18}). Dashed curve 2 is a result of the
evaluation of expression (\ref{2-18}) which is shortened by
dropping out the last square brackets on the r.h.s. of the
equation. Recall that the expression in these square brackets
carries the effect of a relative movement of the pion-pion c.m.s.
and the electron-positron c.m.s., which is a consequence of the
violation of the energy-momentum conservation in s-channel. We see
that, in the region of the electron-positron creation threshold
$M=2m_e$, this effect (Fig.~5, curve 1) is realized as four
additional orders to the creation rate comparing to the rate
without taking into account the relative velocity (curve 2).

\begin{center}
\includegraphics[width=\columnwidth]{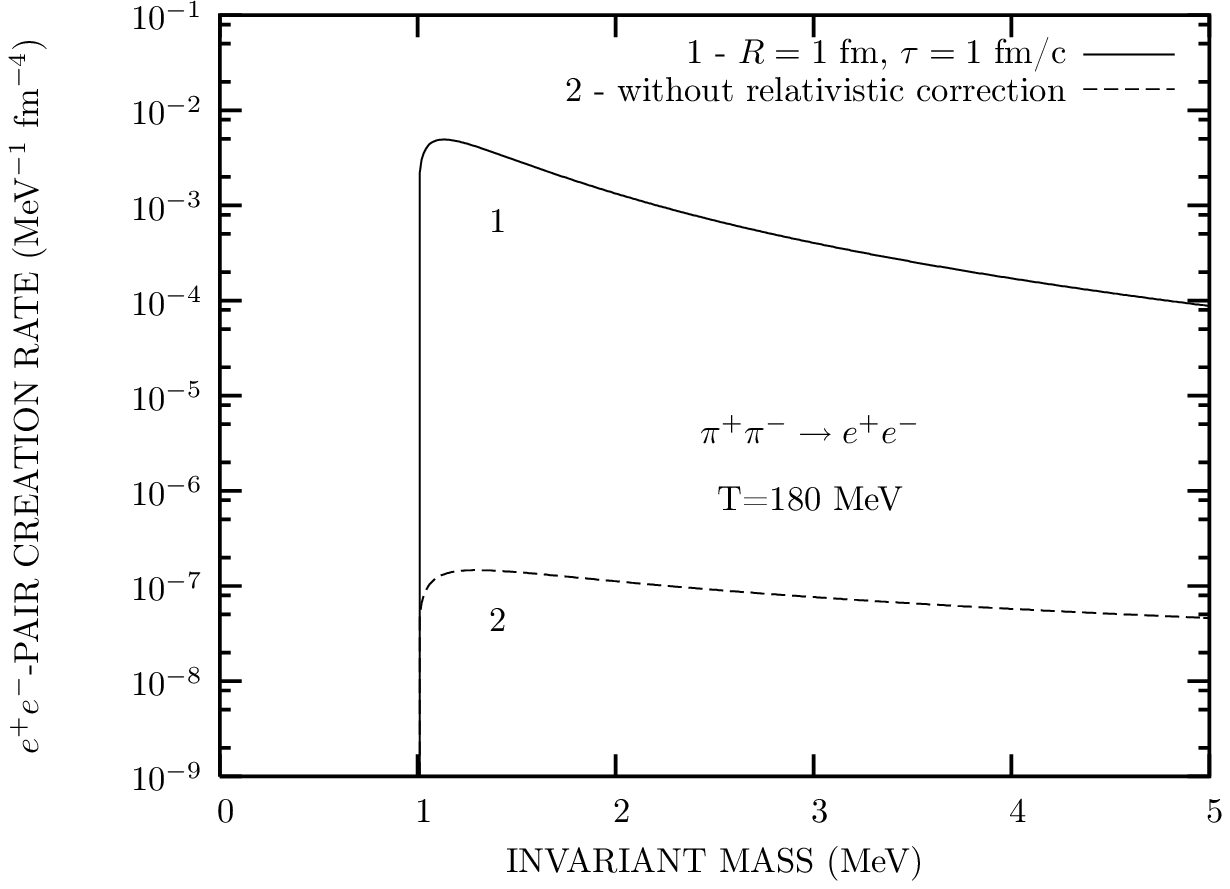}
\end{center}

\noindent {\footnotesize Fig. 5. The rate $dR^{\rm
(\rho)}_{e^+e^-}/dM$ of electron-positron pair creation in
pion-pion annihilation in the threshold region. Mean spatial size
$R=1$~fm, lifetime $\tau=1$~fm/c, $T=180$~MeV. Curve 1 is
evaluated in accordance with expression (\ref{2-18}). Curve 2 is
evaluated for the same integral (\ref{2-18}) but without the last
factor in square brackets which reflects a relative movement of
the pion-pion c.m.s. and the electron-positron c.m.s.
  }
\vspace{5pt}


Fig.~6 shows the same evaluations as the previous one but in a
wider range of invariant masses. Curves 1, 2, 3 are evaluated in
accordance with formula (\ref{2-18}), whereas, curves 1n, 2n, 3n
are evaluated in accordance with Eq.~(\ref{2-18}), where the
relativistic factor [the last square brackets on the r.h.s. of
(\ref{2-18})] is dropped out.

\medskip
\begin{center}
\includegraphics[width=\columnwidth]{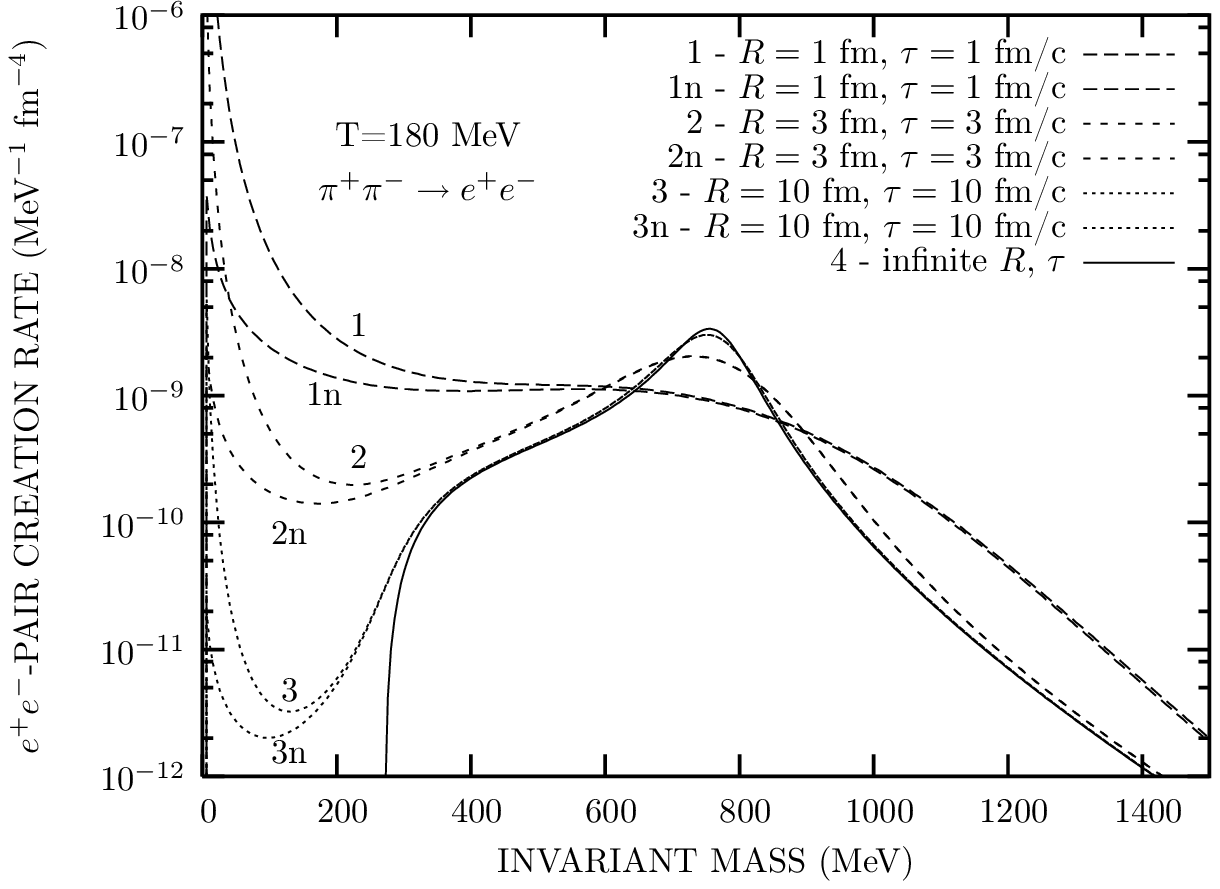}
\end{center}

\noindent {\footnotesize Fig. 6. The rate $dR^{\rm
(\rho)}_{e^+e^-}/dM$ of electron-positron pair creation in
pion-pion annihilation with (curves 1, 2, 3) and without (curves
1n, 2n, 3n) a c.m.s. relativistic correction. Different curves
correspond to the different spatial sizes $R$ and different life
times $\tau$ of the pion system. Curve 4 corresponds to the
infinite size of the pion system.
  }
\vspace{5pt}

\noindent Dash-dotted curves 1 and 1n correspond to $R=1$~fm,
$\tau=1$~fm/c; solid curves 2 and 2n correspond to $R=3$~fm,
$\tau=3$~fm/c; dashed curves 3 and 3n correspond to $R=10$~fm,
$\tau=10$~fm/c, and long-dash-dotted curve 4 corresponds to the
infinite space-time volume of the reaction region. We see that the
relativistic effect under consideration is attributed to the
region of invariant masses which is below the invariant mass
$M=2m_\pi$ and increases with decrease in the size of the pion
system. The latter is partially explained by the presence of last
two terms, $P\cdot K/R^2$ and $1/(2R^4)$, in square brackets on
the r.h.s. of (\ref{2-18}). Notice that this effect for the muon
pair creation rate is practically not visible even for the pion
system of the size $R=1$~fm, $\tau=1$~fm/c, because the real
threshold $2m_\mu$ is close to the $2m_\pi$.

Next, we compare the emission rates obtained in the frame of two
models (a) and (b). Model (a) (see Fig.~2a) reflects the presence
of a dense hadron environment which affects the $\rho$-meson
propagation: due to the medium effects, the $\rho$-meson mean free
path and mean lifetime are strongly suppressed inside the dense
nuclear matter.

\begin{center}
\includegraphics[width=\columnwidth]{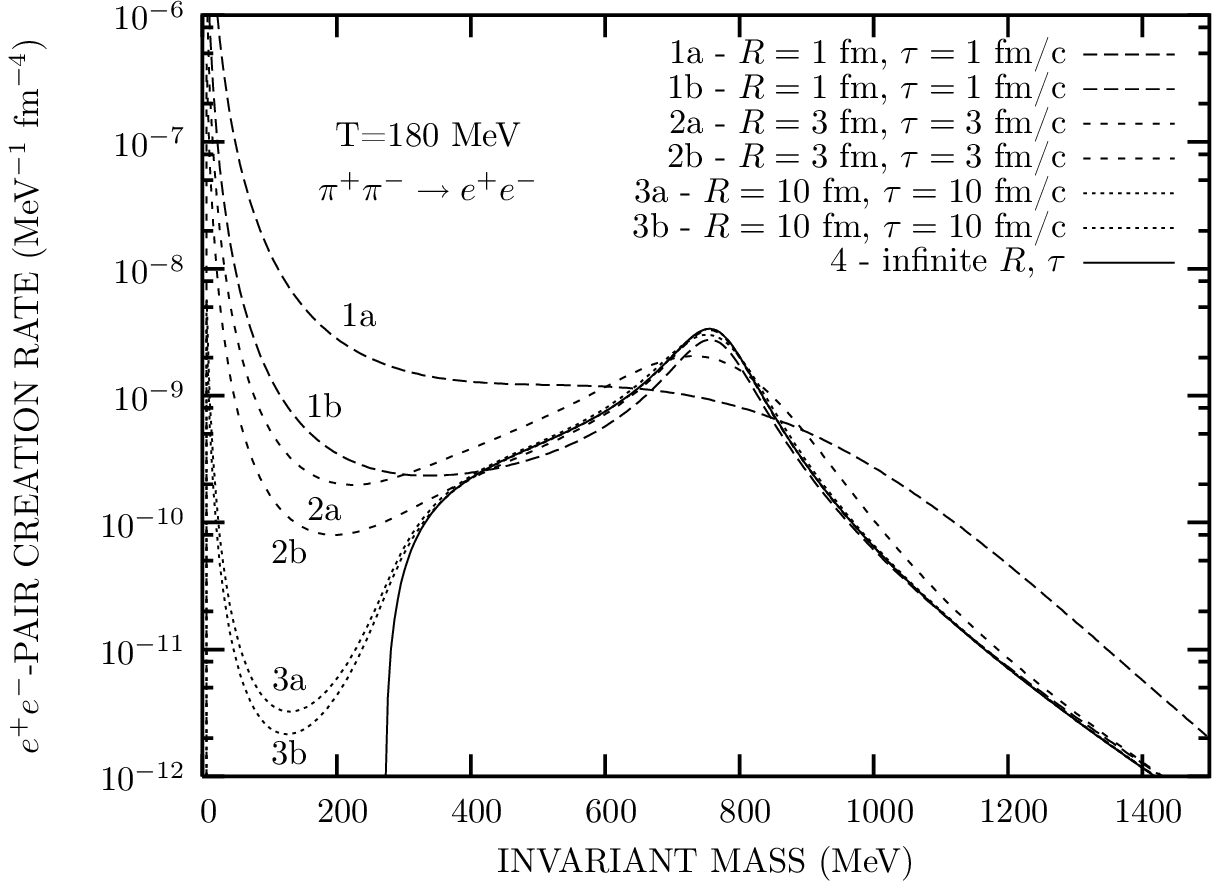}
\end{center}

\noindent {\footnotesize Fig. 7. The rate $dR^{\rm
(\rho)}_{e^+e^-}/dM$ of electron-positron pair creation in
pion-pion annihilation. All curves which are labelled by letter
"a"{} were evaluated in the model which corresponds to the diagram
depicted in Fig.~1a. All curves which are labelled by letter "b"{}
were evaluated in the model which corresponds to the diagram
depicted in Fig.~1b. The values of the mean spatial size and life
time: dash-dotted curves (1a and 1b) - $R=1$~fm, $\tau=1$~fm/c;
dashed curves (2a and 2b) - $R=3$~fm, $\tau=3$~fm/c; dotted curves
(3a and 3b) - $R=10$~fm, $\tau=10$~fm/c; solid curve (4) -
infinite space-time volume. Effective temperature $T=180$~MeV.
  }
\vspace*{10pt}%

\noindent As a limit of this suppression,  we adopt in model (a) a
zero mean free path and zero lifetime keeping vacuum parameters of
the $\rho $-meson form-factor (\ref{2-03}). Model (b) (Fig.~2b) is
a contra-pole to model (a): the $\rho $-meson propagation is not
affected at all by the hadron environment. Then, the real picture
of the pion annihilation in a dense nuclear matter will be
somewhere between these two limit models. Fig.~7 shows the rate
$dR^{\rm (\rho)}_{e^+e^-}/dM$ of electron-positron pair creation
evaluated in the frame of models (a) and (b). Curves 1a, 2a, 3a
are obtained as above by the evaluation of formula (\ref{2-18}).
Curves 1b, 2b, 3b are obtained by the evaluation of formula
(\ref{2-18}), when the argument of the pion form-factor is changed
from $K^2$ to $M^2$. Then it can be written in front of the
integral in the form $\big| F_\pi(M^2) \big|^2$. This change
corresponds to the integration of the vertex $x_1'$ (see Fig.~2b)
over infinite space-time volume. Dash-dotted curves 1a and 1b
correspond to $R=1$~fm, $\tau=1$~fm/c; dashed curves 2a and 2b
correspond to $R=3$~fm, $\tau=3$~fm/c; dotted curves 3a and 3b
correspond to $R=10$~fm, $\tau=10$~fm/c, and solid curve 4
corresponds to the infinite space-time volume of the reaction
region. The temperature of the locally equilibrated pion system
was taken $T=180$~MeV. The comparison of curves 1a and 1b
evidently shows that the $\rho$-meson peak becomes much more
pronounced, when we shift from model (a) to model (b). Hence, we
can conclude that, due to the small mean size and lifetime of the
pion subsystem, we obtain a strong effective smearing of the
$\rho$-meson form-factor $\left| F_\pi(K^2) \right| ^2$ in the
frame of model (a). But this difference becomes smaller with
increase in the mean parameters $R$ and $\tau$. For instance, the
behavior of curves 2a and 2b is qualitatively the same for
$R=3$~fm, $\tau=3$~fm/c.

Fig.~8 shows the emission rate $dR^{\rm (\rho)}_{\mu^+\mu^-}/dM$
of the $\mu^+\, \mu^-$ pair creation for models (a) and (b). We
see that model (a) again reveals indeed a very strong smearing of
the $\rho$-meson peak because of the presence of the pion system
form-factor $|\rho(K-P)|^2$, when it characterized by the smallest
parameters under consideration: $R=1$~fm, $\tau=1$~fm/c.
Meanwhile, for the same set of parameters, the $\mu^+\, \mu^-$
emission rate evaluated in the frame of model (b) (dash-dotted
curve 1b) practically coincides with the standard result (solid
curve 4) in the region $M \geqslant 450$~MeV. But, for invariant
masses $M \leqslant 450$~MeV, the emission rate reveals a strongly
pronounced flat shoulder with a cut off at the threshold
$M=2m_\mu=211.3$~MeV. For the parameters $R=3$~fm, $\tau=3$~fm/c,
the emission rates for models (a) and (b) are not so
distinguishable for invariant masses above the $\rho$-peak (dotted
curves 2a and 2b). But when the invariant mass is below the
$\rho$-peak, these emission rates manifest a well visible corridor
where, as we discussed before, the real rate should be~found.

\begin{center}
\includegraphics[width=\columnwidth]{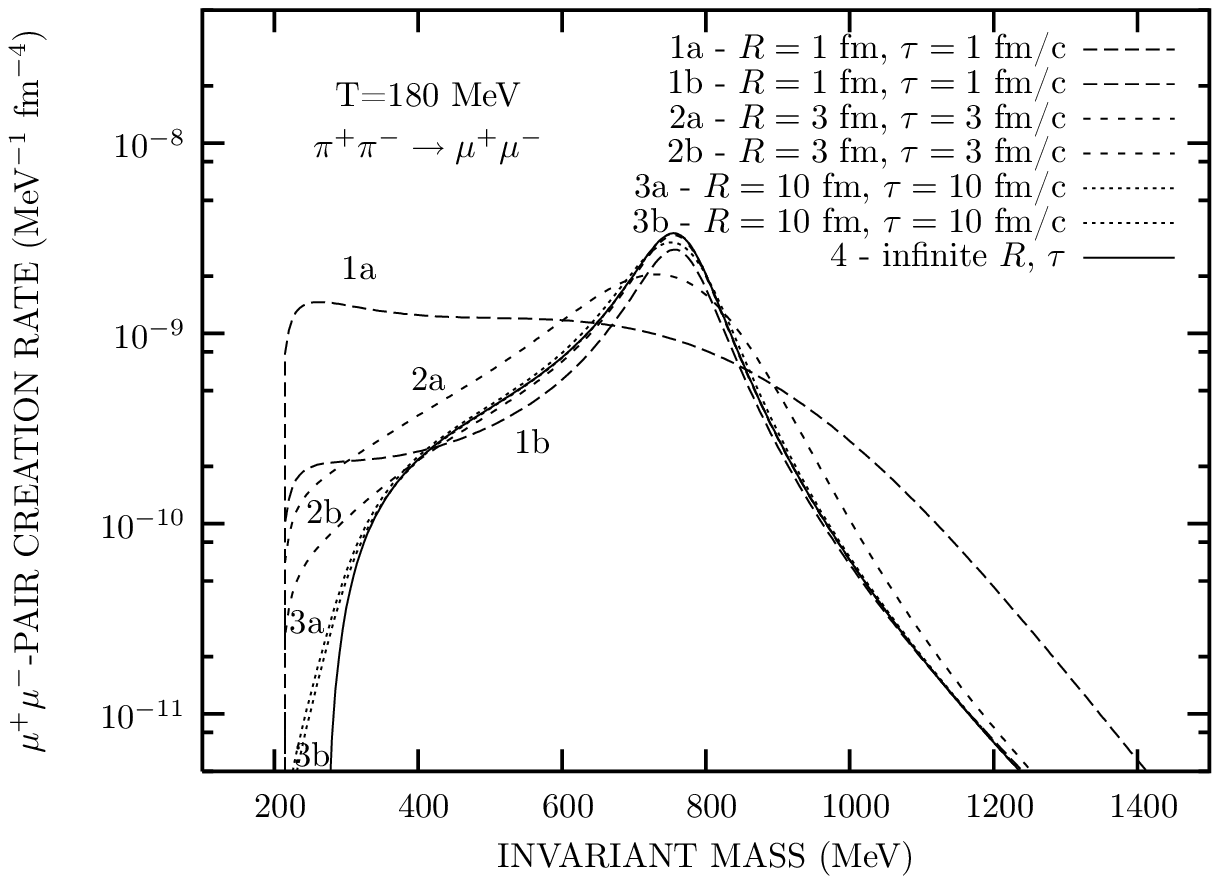}
\end{center}

\noindent {\footnotesize Fig. 8. The rate $dR^{\rm
(\rho)}_{\mu^+\mu^-}/dM$ of $\mu^+\, \mu^-$ pair creation in
pion-pion annihilation. All curves are labelled as in Fig.~7. The
set of parameters is also taken as in Fig.~7.
  }
\vspace*{5pt}%

At last, when the mean size of the pion subsystem is $R=10$~fm,
i.e. is comparable to the size of the fireball, and the mean
lifetime is $\tau=10$~fm/c which is also comparable with the
lifetime of the fireball, the emission rates evaluated in the
frame of models (a) and (b) are not distinguishable (dense dotted
curves 3a and 3b), but a cut off of the rates is at the threshold
$M=2m_\mu$.

For comparison, we present in Figs.~9 and 10 the results of the
evaluation of the rate $dR^{(\rho)}_{e^+e^-}/dM$ of
electron-positron pair creation in a QGP (quark-gluon plasma) drop
from quark-antiquark annihilation.

\medskip

\begin{center}
\includegraphics[width=1.0\columnwidth]{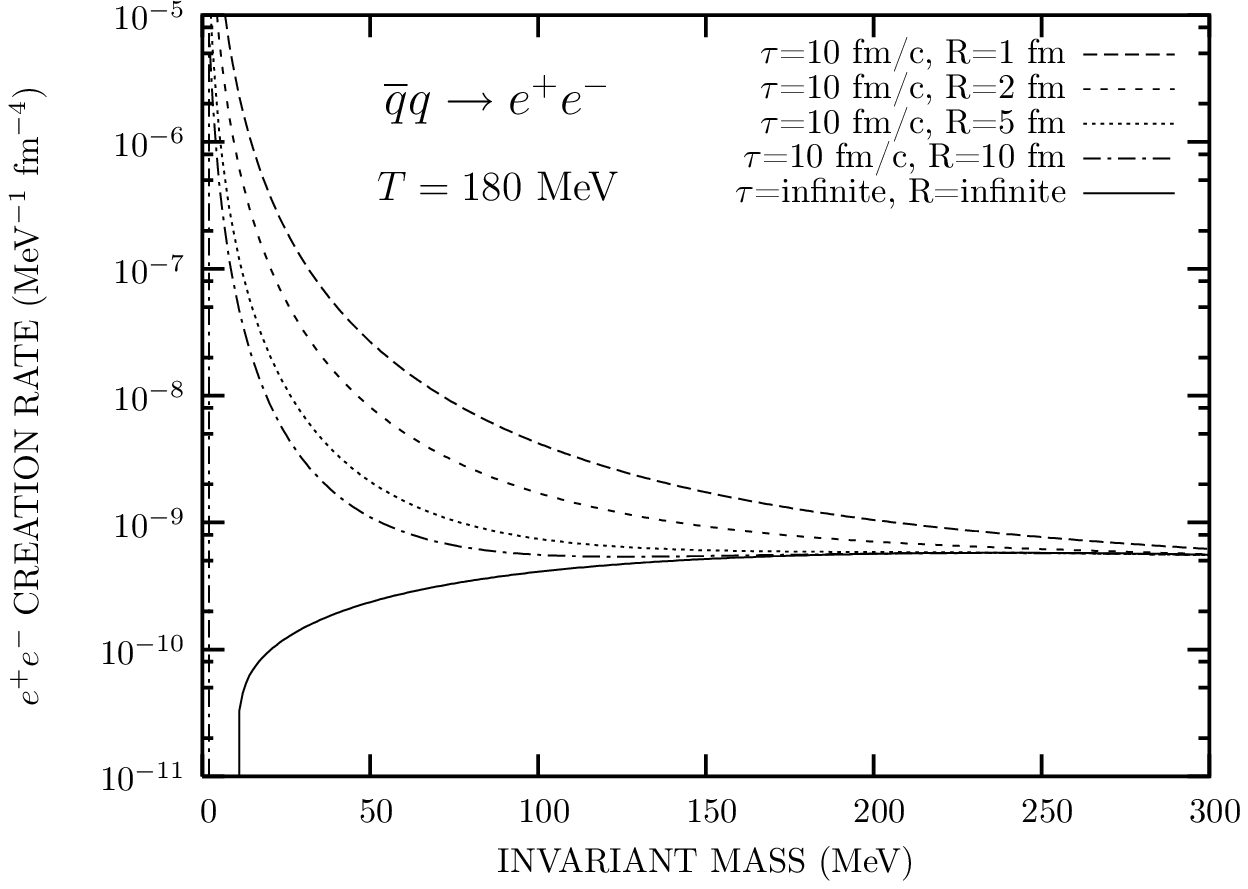}
\end{center}
\vspace*{-10pt}%

\noindent {\footnotesize Fig. 9. The rate $dR^{\rm
(\rho)}_{e^+e^-}/dM$ of electron-positron pair creation in
quark-antiquark annihilation. Different curves correspond to the
different spatial sizes $R$ at the lifetime $\tau=10$~fm/c of a
QGP drop, $T=180$~MeV.
  }


\noindent The evaluation was carried out under the same
assumptions as for pion-pion annihilation. As in the previous
case, we see the increase of the rate with decrease in the
invariant mass up to the two-quark mass threshold, i.e.
$M=2m_q=10$~MeV.

\begin{center}
\includegraphics[width=1.0\columnwidth]{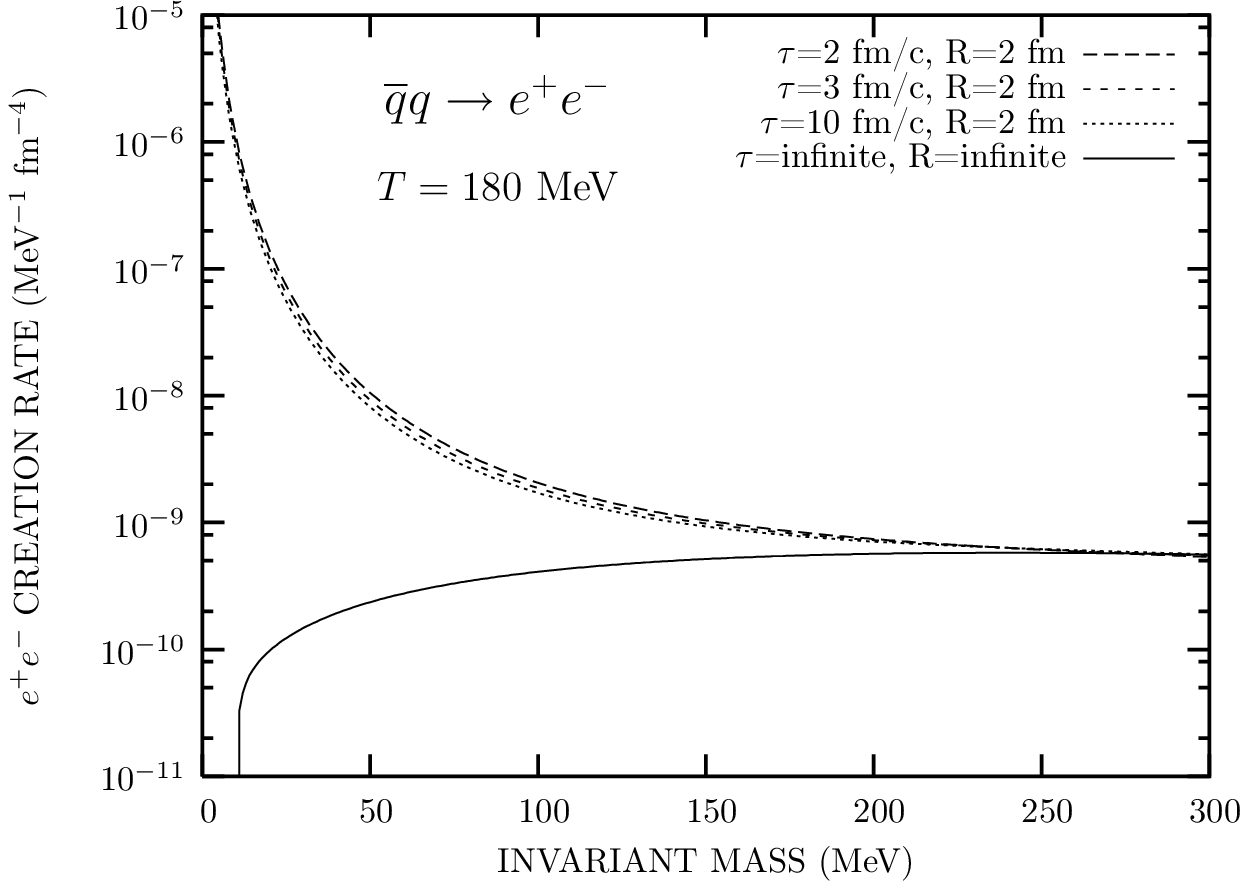}
\end{center}

\medskip

\noindent {\footnotesize Fig. 10. The rate $dR^{\rm
(\rho)}_{e^+e^-}/dM$ of electron-positron pair creation in
quark-antiquark annihilation. Different curves correspond to the
spatial size $R=2$ fm and the different life times $\tau$ of QGP
drop, $T=180$~MeV.
  }
\medskip


\noindent The different behavior of the rate (compared to the case
of infinite space-time) for small invariant masses $M\leqslant
300$ MeV is due to quantum fluctuations which occurs because of a
finite size of the QGP drop. 
It should be pointed out the absence
of the dependence of the rate on the lifetime of a quark-antiquark
system when the size of the system is comparatively small, for
instance $R=2$~fm like in Fig.~10. 
This fact can be explained with
the use of the quark (or pion) density taken in particular form
(\ref{2-14a}). 
Indeed, for the system which is confined in three
spatial dimensions, every of three spatial coordinate densities in
(\ref{2-14a}), for instance, $\exp{(-x_i^2/R^2)}$ with $i=1,2,3$,
corresponds to one integration in (\ref{2-9}) over a component of
the annihilating pair total momentum $K_i$. 
Hence, there are three
integrations on the r.h.s. of (\ref{2-9}) every of each carries a
factor, which is responsible for quantum fluctuations in its own
spatial dimension, for instance, one of three factors is
$(2\pi)^{1/2}R\cdot \exp (-R^2K_i^2)$. 
Meanwhile, the spectral
function, which is responsible for the finiteness of the system
lifetime, for instance $(2\pi)^{1/2}\tau\cdot \exp (-\tau^2K_0^2)$
as in our case, emerges as an integrant factor just once during
integration over $K_0$. 
That is why the variations of the spatial
size of a many-particle system influence the shape of the dilepton
production rate much stronger than the variations of the system
lifetime.

\section{Discussion and conclusions}

In the present work, we have studied the effect of pion-pion and
quark-quark annihilation in a finite space-time volume on the
dilepton yield. Due to the uncertainty principle, a restriction of
the volume and time of these reactions results in breaking the
detailed energy-momentum conservation in the $s$-channel of the
reactions. Formally, this expresses in the alteration of
$(2\pi)^4\delta ^4(K-P)$ to the distribution $\rho(K-P)$ in the
amplitude of the reaction, where $|\rho(K-P)|^2$ appears as the
form-factor of a multipion subsystem (quark-antiquark system),
$K=k_1+k_2$ is the total 4-momentum of an annihilating pion
(quark) pair, and $P=p_++p_-$ is the total 4-momentum of the
created lepton pair. For this kinematics, the energy-momentum
conservation is valid as an integral law which is guaranteed by
the relation $\int [d^4K/(2\pi)^4] \rho(K-P)=1$, where the
registered 4-momentum $P$ plays the role of the center (or the
mean value) of the distribution $\rho(K-P)$ with respect to the
total 4-momentum $K$ of an annihilating pair as a fluctuating
quantity. As a result of the above mechanism, the rate of lepton
pair yield is a finite quantity below the "threshold" \,
$M=2m_\pi$. Actually, the real threshold is determined as the
total mass of the registered particles, i.e., $M_{\rm
thresh}^{e^+e^-}=2m_e$ for $e^+\, e^-$ production (see Fig.~5),
where $m_e$ is the electron mass, and $M_{\rm
thresh}^{\mu^+\mu^-}=2m_\mu$ for  $\mu^+\, \mu^-$ production (see
Fig.~4), where $m_\mu$ is the mass of a $\mu$-meson. In the
calculations made, we kept the invariant mass of a pion-pion
(quark-quark) pair not lesser than $2m_\pi$ ($2m_{\rm q}$), or the
range of integration over the total 4-momentum $K$ of a pion
(quark) pair was determined as $K^2 \geqslant 4m^2_\pi$ ($K^2
\geqslant 4m^2_{\rm q}$).

At first sight, the appearance of finite values of the rate below
the invariant mass $M=2m_\pi$ can look paradoxical. Meanwhile, we
recall that, in the calculation of the rate (see (\ref{2-12}) and
(\ref{2-15a})), replacing a $\delta$-function in the integrand by
a distribution $|\rho(K - P)|^2$ results in the additional
integration where the total momentum of a registered pair $P$ or
its invariant mass $M=\sqrt{P^2} $ plays the role of an external
parameter. Then, if the distribution $|\rho(K - P)|^2$ is wide
enough, it can lead to a finite result of integration for those
values of $P$ or $M$ which gave zero value of the integral in the
presence of the $\delta$-function $\delta^4(K - P)$.

To give the intuitive feeling how it is going on, let us consider
the toy averaging of a "good" \, function $ R(M') $ over the
fluctuating variable $M'$ by using the distribution function $
\rho(M'-M)= (\tau /\sqrt{\pi})\, \exp{[-(M'-M)^2 \tau^2] } $ which
has the mean value $M$ (measurable quantity) and the width of the
distribution $1/\tau$. Then, this averaging reads
\begin{equation}
\langle R\, \rangle (M) =
\int _{2m_\pi}^\infty dM'~R(M')\,
\frac{\tau}{ \sqrt{\pi} } \,
e^{ -\left( M' - M\right)^2 \tau^2 }
\ ,
\label{4a}
\end{equation}
where the variable $M'$ (an analog of the two-pion invariant mass)
varies in the region $ 2m_\pi < M' < \infty $. For long enough
lifetimes ($\tau \to \infty$), we immediately obtain, as a result
of averaging the function $R$ taken at the value $M'=M$, $\langle
R \rangle (M) = R(M) \theta(M-2m_\pi) $. The presence of the
$\theta$-function in this expression stresses the fact that, for
the infinite lifetime, the invariant mass $M$ of a lepton pair
should be bigger than the total mass of annihilating particles, it
is $2m_\pi$ in the case under consideration. Meanwhile, for a
finite lifetime, for instance $\tau=1\div 10$~fm/c, the integral
is not zero for mean values of the distribution $M$ which are
lesser than $2m_\pi$. For example, we can consider even the
extremely small value $M=2 m_e \approx 1$~MeV. Then, though the
distribution function $\rho(M'-M)$ is centered around the mean
value $M=2 m_e$, i.e. beyond the region of integration $ 2m_\pi
\leqslant M' < \infty $, the right wing of the distribution
$\rho(M'-M)$ is valued enough to make the result of integration on
the r.h.s. of (\ref{4a}) quite significant to be taken into
account. The illustration of this evaluation is depicted in
Fig.~11, where the integrand in (\ref{4a}) is shown schematically
as a solid line. The value of integral (\ref{4a}) $\langle R\,
\rangle (M)$ equals the shaded area which starts from the value
$M'=2m_\pi$.

%
\begin{center}
\includegraphics[width=0.8\columnwidth]{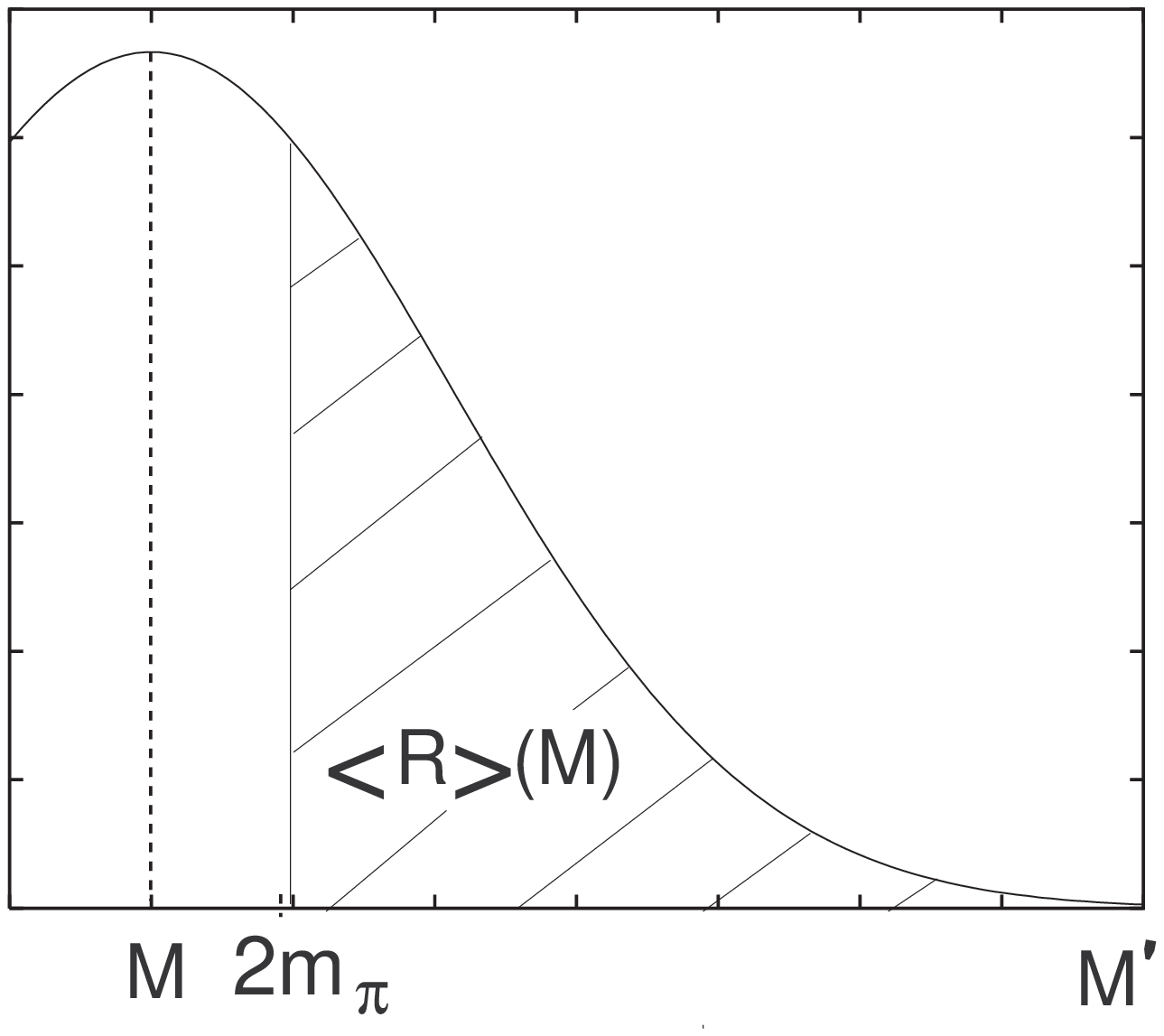}
\end{center}
\vspace{-10pt}

\noindent {\footnotesize Fig. 11. Sketch of the integrand in
Eq.~(\ref{4a}). The integral is evaluated when the value of the
parameter $M$ is lesser than the lower limit of integration,
$2m_\pi$.
  }

\vspace*{6pt}%

\noindent In a similar way, if the lifetime or/and size of the
multipion system is small enough, we obtain a nonvanishing rate
for dilepton invariant masses which are lesser than $2m_\pi$,
whereas the invariant masses of the two-pion system (it is $M'$ in
our toy example) is certainly bigger than $2m_\pi$.

One of the consequences of breaking the exact equality between the
total 4-momentum $K$ of an incoming pion pair and the total
4-momentum $P$ of an outgoing lepton pair is a noncoincidence of
two center-of-mass systems: velocity of the $\pi^+\, \pi^-$
c.m.s., for instance in lab system, differs from velocity of the
$l^+\, l^-$ c.m.s. This results in the following: if some quantity
was obtained through calculations in the $l^+\, l^-$ c.m.s.  for
its use in the $\pi^+\, \pi^-$ c.m.s. this quantity should be
Lorentz-transformed. The Lorentz transformation brings an
additional deviation of the dilepton yield from the standard one.
As seen from Figs.~5,~6, the manifestation of this effect for
$e^+\, e^-$ production is essential in the region which is below
$M=2m_\pi$ and the effect is washed out practically for invariant
masses which are bigger than two pion masses. Note that, this
effect is not visible practically for $\mu^+\, \mu^-$ production
(see Fig.~4) because the threshold invariant mass is big enough,
$M_{\rm thresh}^{\mu^+\mu^-}=2m_\mu \approx 211.3$~MeV, and lies
very close to $M=2m_\pi$. We use the concept that a dense hadron
medium decreases the $\rho$-meson mean free path and lifetime. As
a limit of this diminution, one can assume the zero mean free path
and the zero mean lifetime of a $\rho$-meson. We considered two
models of pion-pion annihilation in a hadron medium. The first
model (a) (Fig.~2a) mimics an extreme influence of the dense
hadron environment through adopting a $\rho$-meson mean free path
and mean lifetime to be equal to zero (a $\rho$-meson does not
propagate in the hadron environment), whereas the parameters of
the $\rho$-meson form-factor is taken as the vacuum one. In
accordance with the second model (b) (Fig.~2b), a $\rho$-meson
propagates in the hadron environment without any suppression due
to the environment (the vacuum parameters of the $\rho$-meson
propagator are adopted as well). The models under consideration
reflect two limiting cases: an extreme influence of the dense
hadron medium [model (a)] and the absence of influence of the
hadron environment [model (b)] on the $\rho$-meson mean free path
and mean lifetime. Then, the evaluations of dilepton rates in the
frame of both models mean the determination of the corridor which
should absorb a real rate. We compared the rates evaluated in the
frame of both models and found that the difference between them is
considerable just for small sizes of the multipion systems (see
Figs.~7,~8). So, if annihilating (interacting) particles are
confined to a finite space-time region, this inspires several
effects which give rise to the new features of the dilepton
production rate in comparison to the standard rate, which is
attributed to the infinite space-time. In the present paper, we
considered just two of these effects: 1)~Direct contribution from
the multiparticle system form-factor and 2) The relativistic
correction which is due to the difference of velocities of the
c.m.s. of outgoing particles ($l^+\, l^-$ pair) and the c.m.s. of
incoming particles ($\pi^+\, \pi^-$ or $q \overline{q}$ pair). It is
found that even these two contributions give a significant effect
on the dilepton yield. Meanwhile, the third contribution which
comes from the nonstationarity of a confined multiparticle
(multipion, multiquark) system is still left beyond the scope of
the paper. We reserve a consideration of this problem to the next
paper which is in progress as well as the application of the
approach elaborated to other reactions which take place in
quark-gluon and hadron plasmas.

\section*{Acknowledgements}

One of the authors (D.A.) would like to express his gratitude for warm
hospitality to the staff of the Physics Department of the University
of Jyv\"askyl\"a where this work was started.
D.A. is thankful to  U.~Heinz,  J.~Pisut, E.~Suhonen and V.~Malnev for
discussions and support.


\end{multicols}

\end{document}